\documentclass[letterpaper,useAMS,usenatbib]{mn2e}
\usepackage{amssymb,amsmath,graphicx,lscape,upgreek,float,rotating,rotate,longtable,caption}
\usepackage[absolute]{textpos}
\usepackage{aas_macros}
\usepackage[T1]{fontenc}
\usepackage{aecompl}

\usepackage{color}
\usepackage[usenames,dvipsnames]{xcolor}
\usepackage{xspace}
\newcommand{\kms}{{\rm km\,s}^{-1}}
\usepackage{lipsum}

\title[Caught in the act: Cluster `k+a' galaxies as a link between spirals and S0s]{Caught in the act: Cluster `k+a' galaxies as a link between spirals and S0s}

\author[Rodr\'{i}guez Del Pino et al.]{Bruno Rodr\'{i}guez Del Pino,$^{1}$\thanks{E-mail:
ppxbr@nottingham.ac.uk} Steven P. Bamford,$^{1}$ Alfonso
Arag\'{o}n-Salamanca,$^{1}$\newauthor
Bo Milvang-Jensen$^{2}$, Michael R. Merrifield$^{1}$ and Marc Balcells$^{3,}$$^{4,}$$^{5}$
\smallskip\\
$^{1}$ School of Physics and Astronomy, The University of Nottingham,
University Park, Nottingham, NG7 2RD, UK\\
$^{2}$ Dark Cosmology Centre, Niels Bohr Institute,
University of Copenhagen,
Juliane Maries Vej 30, 2100 Copenhagen {\O}, Denmark\\
$^{3}$ Isaac Newton Group of Telescopes, Apartado 321, E-38700 Santa Cruz de La Palma, Canary Islands, Spain\\
$^{4}$ Instituto de Astrof\'{\i}sica de Canarias, E-38200 La Laguna, Tenerife, Spain\\
$^{5}$ Dep. de Astrof\'{\i}sica, Universidad de La Laguna, E-38205 La Laguna, Tenerife, Spain
}

\begin{document}


\pagerange{\pageref{firstpage}--\pageref{lastpage}} \pubyear{2013}

\maketitle

\label{firstpage}

\begin{abstract}
  We use integral field spectroscopy of 13 disk galaxies in the
  cluster AC114 at $z \sim 0.31$ in an attempt to disentangle the
  physical processes responsible for the transformation of spiral
  galaxies in clusters. Our sample is selected to display a dominant
  young stellar population, as indicated by strong H$\delta$
  absorption lines in their integrated spectra.  Most of our galaxies
  lack the [OII]$\lambda$3727 emission line, and hence ongoing star
  formation.  They therefore possess `k+a' spectra, indicative of a
  recent truncation of star formation, possibly preceded by a
  starburst. Disky `k+a' galaxies are a promising candidate for the
  intermediate stage of the transformation from star-forming spiral
  galaxies to passive S0s.

  Our observations allow us to study the spatial distributions and the
  kinematics of the different stellar populations within the
  galaxies. We used three different indicators to evaluate the
  presence of a young population: the equivalent width of H$\delta$,
  the luminosity-weighted fraction of A stars, and the fraction of the
  galaxy light attributable to simple stellar populations with ages
  between $0.5$ and $1.5$ Gyr.  We find a mixture of behaviours, but
  are able to show that in most of galaxies the last episode of
  star-formation occured in an extended disk, similar to preceeding
  generations of stars, though somewhat more centrally concentrated.
  We thus exclude nuclear starbursts and violent gravitational
  interactions as causes of the star formation truncation.  Gentler
  mechanisms, such as ram-pressure stripping or weak galaxy-galaxy
  interactions, appear to be responsible for ending star-formation in
  these intermediate-redshift cluster disk galaxies.
\end{abstract}


\section{Introduction}\label{sec:introduction}

The properties of galaxies -- such as morphology, 
colour, size and mass -- vary according to the environment 
where they reside. In 
particular, galaxy morphologies have been shown to change 
with local projected density \citep[e.g.,][]{Dressler_1980,Bamford_2009}, with 
late-type spiral and irregular galaxies showing more 
preference for regions with lower densities, while early-type 
S0 and elliptical galaxies are more abundant in denser regions. 
Although in a different timescale, specific star formation rates, 
SSFR, are also affected by the environment \citep{Balogh_2004,Vogt_2004b}
and has been shown to be the galaxy property most affected 
by the density of the environment \citep{Kauffman_2004, Wolf_2009}. 
The concentration of the star formation in cluster disk galaxies is also found
to be $\sim$ 25 per cent smaller than comparable galaxies in the field \citep{Bamford_2007}.

There is also a change in the morphological make-up of the galaxy
population with redshift, particularly in clusters.  Spiral galaxies
show high fractions in clusters at intermediate redshift (z $\sim$ 0.5), where the
fraction of S0s is low, but while the fraction of spirals decreases
for local clusters, S0s become more dominant, being 2--3 times more
abundant today than at intermediate redshift \citep{Dressler_1997}. On
the other hand, ellipticals do not show a substantial variation,
comprising a significant fraction of cluster galaxies since at least $z \sim
1$ \citep{Dressler_1997, Fasano_2000, Desai_2007}.  Correspondingly,
the fraction of star-forming blue galaxies in clusters has been shown
to increase with redshift \citep{BO_1978,BO_1984,Margoniner_2001},
known as the Butcher-Oemler effect, and these have been found to
comprise normal late-type spirals \citep{Dressler_1994, Couch_1994}.

All these different findings point to a transformation of galaxies
from spiral into S0 within the cluster environment, as suggested in
many studies \citep[e.g.,][]{Larson_1980, Shioya_2002, Bekki_2002,
  Alfonso_2006}. This transformation would start with blue,
star-forming, spiral galaxies at intermediate redshift falling into
regions of higher density such as groups and clusters, experiencing
the loss of their gas and subsequent suppression of star formation,
but retaining their disks, resulting in red, passive, S0 galaxies.

A variety of mechanisms have been suggested to be responsible for such
transformations: interaction with the hot intracluster medium (ICM) via thermal evaporation \citep{Nipoti_2007}
and ram-pressure stripping \citep{Gunn_Gott_1972, Abadi_1999, Bekki_2002},
 interactions with the cluster tidal field \citep{Larson_1980}, 
galaxy harassment \citep{Moore_1996} and minor mergers \citep{Bekki_2005, Eliche_2012, Eliche_2013}.
 All these processes may be expected to remove or
disturb the gas contents of galaxies, while leaving the stellar
distributions relatively unscathed. Major mergers can also
trigger starbursts which may consume gas reservoirs and ultimately
supress star-formation \citep{Mihos_1996}, although unless they are
gas-rich \citep{Hopkins_2009}, their stellar disks may be disrupted.
Importantly, none of these mechanisms is thought to operate equally 
from low-mass groups to rich clusters.  The high fraction of S0s present in all these dense
environments therefore suggests that a combination of these mechanisms
may be involved, with varying degrees of importance.

Galaxies in which star formation has been recently suppressed, $\sim
0.5$--$1.5$ Gyr ago, should be well described by the composite of an
A-type stellar population (whose main-sequence lifetime ranges from
$0.5$--$1.5$ Gyr) and an old population, which was present in the
galaxy before the last episode of star formation. This type of
galaxies was found for the first time by \citet{Dressler_Gunn_1983}, and
they are conspicuous by the presence of strong Balmer absorption
lines in their spectra, characteristic of the A stars, superimposed
onto a spectrum of an older (several Gyr) stellar population, and with no emission lines
(indicating no ongoing star formation). These galaxies are called
either `k+a', after their dominant stellar types (old `k' and young
`a'), or `E+A', indicating their spectra correspond to that of a
typical early-type (`E') galaxy with additional A-stars. We will refer
to them as `k+a' galaxies hereafter.

Due to the importance of `k+a' galaxies as observable instances of
rapid evolution, they have been the subject of many studies
\citep{Dressler_Gunn_1983, Zabludoff_1996, Norton_2001, Pracy_2009,
  Poggianti_2009a, Pracy_2012, Pracy_2013}. Although first discovered in the
cluster environment \citep{Dressler_Gunn_1983}, they have also been
found in the field \citep{Zabludoff_1996, Blake_2004} and in groups
\citep{Poggianti_2009a}. Few `k+a' galaxies are found in the local universe, 
but their prevalence increases significantly with redshift, such that 
in intermediate-redshift clusters they can represent up to 10 per cent 
of the total galaxy population \citep{Poggianti_2009a}.In those intermediate-redshift clusters,
`k+a' galaxies tend to avoid the central regions, implying that the
suppression of star formation does not require the extreme conditions
of cluster cores, and may begin in less dense environments such as
groups \citep{Dressler_1999}. While `k+a' galaxies in general often
show early-type morphologies (sometimes disturbed; \citealt{Yang_2008}),
in clusters they are generally associated with disk-like systems
\citep{Caldwell_1999, Tran_2003}, and in many cases they also show
spiral signatures, implying that the timescale for the spectral
evolution is shorter than that for any morphological transformation
\citep{Poggianti_1999}.

Analysing the internal spatial
distributions, and ideally kinematics, of the different stellar
populations inhabiting these galaxies is crucial to understanding the
mechanisms responsible for the suppression of star formation. If the
last episode of star formation took place in the central regions, it
would be consistent with processes such as galaxy-galaxy interactions
and minor mergers (\citealt{Mihos_1996, Bekki_2005}, but see, e.g.,
\citealt{Teyssier_2010}, and the discussion in Section \ref{section:discussion} in this
paper). In contrast, a more extended young population could
imply depletion of a galaxy's gas reservoir through interaction with
the hot ICM \citep{Rose_2001, Bekki_2005, Bekki_2009}.

To perform such an analysis, we have used integral field spectroscopy, obtained using the
FLAMES-GIRAFFE multi-object 
spectrograph at the VLT \citep{Pasquini_2002}, to analyse 13 galaxies with disk morphologies
and strong H$\delta$ absorption in the cluster 
AC114 (also known as Abell S1077; \citet{Abell_1989}) at $z \sim
0.3$.  AC114 has 
been shown to contain a significant population of blue star-forming galaxies 
by \citet[hereafter CS87]{Couch_Sharples_1987}, but also to have a substantial general suppression of the star formation 
(as inferred from H$\alpha$ emission; \citealt{Couch_2001}), which 
makes it an ideal laboratory for studying how cluster galaxies are transformed. 
A previous study of `k+a' galaxies in this cluster has been carried out 
by \citet[hereafter P05]{Pracy_2005}. They obtained observations 
using FLAMES with a very similar configuration, although they did not focus 
specifically on galaxies with disk morphology.  We were not aware that
their observations existed when ours were scheduled, but such repeated 
observations enable us to
check the reproducibility of our measurements.  Combining the P05
dataset with our own also adds some additional galaxies to the sample
we consider in this paper. In their study, P05 only consider 
the spatial distribution of the H$\delta$ equivalent width. We expand on this, measuring
the stellar populations in more detail and considering the resolved galaxy kinematics. 

Throughout this paper we assume a cosmology with $\Omega_{\rm m} = 0.3$, 
$\Omega_{\Lambda} = 0.7$, 
$H_{\rm 0} = 70 \,\kms$Mpc$^{-1}$.


\section{Data}

\subsection{Sample}

The current sample consists of 13 galaxies, observed and identified by CS87 as
members of the cluster AC114 at $z \sim 0.31$.  The CS87 catalogue
provides redshifts and spectral line measurements, as measured on 8
hour integrations with the $3.9$ m AAT, using a spectrograph with a
spectral resolution $R \sim 1400$ and fed by $2.6 \arcsec$ diameter
fibres.  AC114 also has wide-field archive \emph{HST} WFPC2 imaging,
which is used to catalogue the morphological make-up of AC114 in
\citet[hereafter C98]{Couch_1998}.  Based on the combined CS87/C98
catalogue, the sample galaxies were selected to have $\rmn{H}\delta$
rest-frame equivalent width 
$\rmn{EW}(\rmn{H}\delta) > 3$\AA{ }(the sign convention here is that a positive EW for 
 $\rmn{H}\delta$ means absorption), which is the common criterion to be
considered `k+a' \citep{Poggianti_1999}, disk morphology and magnitude
$R_{\rm F} \le 20.5$.  Some further sample limitations were imposed by the
spectrograph's field-of-view and restrictions on the placement of each
integral field unit (IFU) in order to avoid button collisions and
crossed fibres.

Nine of the objects selected to be observed show no [OII]$\lambda 3727$
emission in the CS87 catalogue, and hence correspond to a true `k+a'
selection. The remaining four show some [OII]$\lambda 3727$ emission,
indicating that they have ongoing star formation, though possibly
declining given their H$\delta$ EW, or host an AGN.

In addition to these galaxies, we include in our analysis the
objects observed by P05.  This sample was selected from the same
cluster in a similar manner to that described above, except that no
restriction was placed on morphology and none of their galaxies had 
detected [OII]$\lambda 3727$ emission. Six galaxies from our selection
were also observed by P05, as well as two additional disk galaxies,
four ellipticals and one peculiar galaxy.  The combined sample therefore
comprises twenty galaxies, of which fifteen possess disk morphology.

Note that, for galaxy CN849, the flux present in our observations was
very low.  This was found to be due to an incorrect target
position. Fortunately, this galaxy was also included in the P05 data
sample and could be analysed using that data.  Also, when comparing
the redshifts measured for the galaxies observed by both P05 and
ourselves we discovered an inconsistency for CN254. Inspecting the
coordinates we discovered that the galaxy labelled CN254 in P05 is
actually CN229, another disk galaxy at $z=0.319$.  The cross-comparison
sub-sample with multiple observations therefore comprises four objects.

In Table \ref{table:1} we list all of the objects considered in this
paper, with their coordinates, morphologies,
colour ($B_J$-$R_F$, corrected for Galactic reddening) and projected
distance to the cluster centre.

\subsection{Observations}

The observations were obtained at the VLT-UT2 using the Fibre Large 
Array Multi Element Spectrograph (FLAMES) in GIRAFFE mode at a 
resolution of R $\sim 9600$. With this setup, 15 individual IFUs 
were deployed over the whole field of view, with two of them being 
dedicated to the sky to ensure a reliable sky subtraction. 
Each IFU consists of 20 squared microlenses of $0.52$ arcsec on a side, making up 
a surface of $3 \times 2$~arcsec$^2$ per IFU, which corresponds to 
$\sim 14.0 \times 9.3$~kpc$^2$ at the distance of AC114 ($\sim 2.3 \times 2.3$~kpc$^2$ per spaxel).

The total exposure time was $\sim 13$ hours, distributed in 14 
exposures in different nights of June, August and December of 2004. 
Observations were taken with seeing conditions within the requested service
mode constrain ($\leq$ 0.8 arcsec), and DIMM seeing ranged from 0.49 to 1.06 arcsec.
The observed wavelength range was $5015$--$5831$\AA{}, which at a redshift 
of $z \sim 0.3$ corresponds to $3850$--$4394$\AA{} in rest frame, covering the K and H calcium features 
($3934$\AA{}  and $3969$\AA), the Balmer lines H$\delta$ ($4102$\AA{}) 
and H$\gamma$ ($4341$\AA) and the G-band ($4305$\AA{}). At that wavelength range, the instrumental 
resolution is $0.57$\AA{} sampled with $0.2$\AA{} pixels, yielding 
 a velocity resolution of $\sigma = 10 \,\kms$ at $z 
\sim 0.3$. Since we expect $\sigma \geq 50 \,\kms$, 
this resolution is enough to comfortably resolve the lines.

In order to ensure an accurate calibration of the dataset, we obtained arc lamp and
Nasmyth flatfield images immediately after each science exposure.

The observations by P05 were obtained with an identical setup, though with slightly
lower integration times. The seeing values for
these observations ranged from 0.54 to 0.84 arcsec. Their 
independent spectra for the four galaxies we have in common provide a useful check 
of the robustness of our results.

\subsection{Data reduction}

The data were reduced using the GIRAFFE pipeline
provided by ESO \citep{Izzo_2004}.  The
pipeline first subtracts the bias and the overscan regions.  Then,
using the corresponding Nasmyth flatfield image, it determines the
position and width of the spectra on the CCD and simultaneously
produces a normalised flatfield to account for the variations in
transmission from fibre to fibre. Because the observations were taken
with the original CCD, which was only changed in May 2008, removal of the
dark was necessary due to the presence of a prominent glow in the
CCD. A dispersion solution was created using the corresponding ThAr
arc lamp frame, and the spectra rebinned to a constant dispersion.
No flux calibration was required for the analysis of the data. 

The pipeline did not include a recipe for the subtraction of the
sky. Therefore, the subtraction was done combining all fibres from the
two IFUs dedicated to the sky, together with the single sky fibre
associated with each IFU, giving a total of 52 fibres. We noticed that
one of the sky IFUs was systematically too bright, perhaps due to
contamination by a low surface brightness object, resulting in an
oversubtraction of the sky in our object fibres. We decided to exclude
this IFU and use the remaining 32 sky fibres for the sky subtraction.

For consistency, we obtained the raw data for the P05 observations
from the ESO archive and reduced them in the same manner as our own
observations.

\begin{table}
{\scriptsize
\begin{center}
\smallskip
\begin{tabular*}{0.49\textwidth}{lcccccc}
\hline
\hline
\noalign{\smallskip}

Target  	& $\alpha_{J2000}$ 	& $\delta_{J2000}$	& Morphology  	 	&  $B_{\rm J}-R_{\rm F}$      & $R_{\rm cl}$ \\
		& $\rm h\,m\,s$        	& $^{\circ}$\,$\arcmin$\,$\arcsec$ &       	 	     & 		        & $(Mpc)$  \\
\hline                                                                   
CN24  		& 22\,58\,50.0 		& -34\,47\,57		& Disk 	 		& $2.4$ & $0.10$\\
CN74$^{*}$            & 22\,58\,44.6 		& -34\,49\,11		& Disk 	 			& $1.26$ & $0.34$\\
CN119 		& 22\,58\,42.3 		& -34\,47\,41		& Disk 	 		& $2.44$ & $0.48$\\
CN143$^{P05}$ 		& 22\,58\,42.8 		& -34\,48\,31		& Disk 	 		& $1.67$ & $0.32$\\
CN146$^{*}$          & 22\,58\,49.7 		& -34\,52\,13		& Disk 	 			& $1.42$ & $0.34$\\
CN155$^{*}$	        & 22\,58\,43.4 		& -34\,49\,37		& Disk 	 		& $1.43$ & $0.48$\\
CN187$^{P05}$ 		& 22\,58\,50.5 		& -34\,49\,12		& Disk 	 		& $2.17$ & $0.31$\\
CN191$^{P05}$ 		& 22\,58\,52.9 		& -34\,48\,46		& Disk 	 		& $1.49$ & $0.30$\\
CN228$^{P05}$ 		& 22\,58\,46.5 		& -34\,46\,18		& Disk 	 		& $1.39$ & $0.50$\\
CN232 		& 22\,58\,59.5 		& -34\,51\,46		& Disk 	 		& $2.34$ & $0.50$\\
CN243$^{*}$          & 22\,58\,40.7 		& -34\,46\,44		& Disk 	 			& $1.58$ & $0.56$\\
CN254		& 22\,58\,39.8 		& -34\,47\,50		& Disk 	 		& $1.99$ & $0.48$\\
CN849$^{P05}$		& 22\,58\,37.3s 	& -34\,48\,20		& Disk 	 		& $1.8$ & $0.61$\\
\hline           
CN4 		& 22\,58\,40.7s 	& -34\,47\,53		& Elliptical 		& $2.48$ & $0.42$\\
CN22 		& 22\,58\,50.0s 	& -34\,48\,13		& Peculiar  		& $1.48$ & $0.09$\\
CN89 		& 22\,58\,48.9s 	& -34\,46\,57		& Elliptical 		& $2.24$ & $0.32$\\
CN229 		& 22\,58\,45.3s 	& -34\,46\,21		& Disk 	 		& $2.3$ & $0.51$\\
CN247 		& 22\,58\,39.6s 	& -34\,47\,15		& Elliptical 		& $2.46$ & $0.53$\\
CN667 		& 22\,58\,41.0s 	& -34\,46\,21		& Disk 	 		& $1.6$ & $0.62$\\

CN858 		& 22\,58\,48.0s 	& -34\,47\,26		& Elliptical 		& $2.35$ & $0.19$\\
\hline\hline                                                                                                              
\end{tabular*} 

\caption{Identity number of the galaxy from \citet{Couch_1984}, coordinates, morphologies, Galactic reddening corrected colours and distance to the cluster centre for the galaxies in
our sample (top) and the P05 sample (bottom). At $z = 0.3$ one arc
second corresponds to $4.454$ kpc. Galaxies labeled with * have detected emission in [OII]$\lambda 3727$. Galaxies labeled with "P05" are present in both samples.}\label{table:1} 
\end{center}
}
\end{table}

\subsection{Stellar population and kinematic analysis}\label{sec:ppxf_analysis}

To extract information about the kinematics and stellar populations of the galaxies, we used the penalized 
pixel fitting {\tt pPXF} software described in \citet{Cappellari_Emsellem_2004}. 
This algorithm uses a maximum-likelihood 
approach to fit the spectra in pixel space, simultaneously determining
both the stellar kinematics and the 
optimal linear combination of spectral templates required to match the input 
spectrum.  We employed two separate collections of templates, one
drawn from
the ELODIE 3.1 stellar library \citep{Prugniel_2007} and the other containing
PEGASE-HR simple stellar population (SSP) models
\citep{LeBorgne_2004}.  The latter spectra are constructed using the
ELODIE library, and hence both have the 
high resolution and wavelength coverage required to fit our spectra 
($0.5$\AA{} FWHM and $4000$--$6000$\AA{}, respectively). 
Internally, {\tt pPXF} convolved the template spectra with a Gaussian in 
order to match the spectral resolution of our observations.
We restricted the templates to two classes (II-III and V) for each stellar
type OBAFGKM, and to SSPs with 12 different ages logarithmically distributed between 1Myr and 15 Gyr
and 5 different metallicities [Fe/H] ranging from -1.7 to 0.4.  

For each spectrum, the program outputs the velocity, $V$, and velocity
dispersion, $\sigma$, together with a refined estimate of the
redshift. The values obtained
for the kinematics when using the stellar library templates and those
obtained using the SSP models were, in general, very similar. However,
for some of the galaxy spectra, occasional noise features present in
the stellar library templates spectra resulted in obviously discrepant
fits and wrong values for the kinematics.  In these cases, we only use
the results obtained using the SSP models.  Errors in the kinematic
parameters obtained with {\tt pPXF} were estimated in the recommended
manner, by performing Monte Carlo simulations on the original spectra
with added noise.

In addition to the kinematics, {\tt pPXF} also provides the 
weights of the templates which provide the best fit to the observed 
spectrum.  These weights, after normalisation, represent the
fractional contribution of each
template to the total luminosity. Below we use the weights obtained 
using the stellar library and the SSP models separately, in order to study the 
distribution of different stellar types and stellar populations 
throughout our sample galaxies.

\begin{figure*}
\begin{center}
   \includegraphics[width=0.45\textwidth]{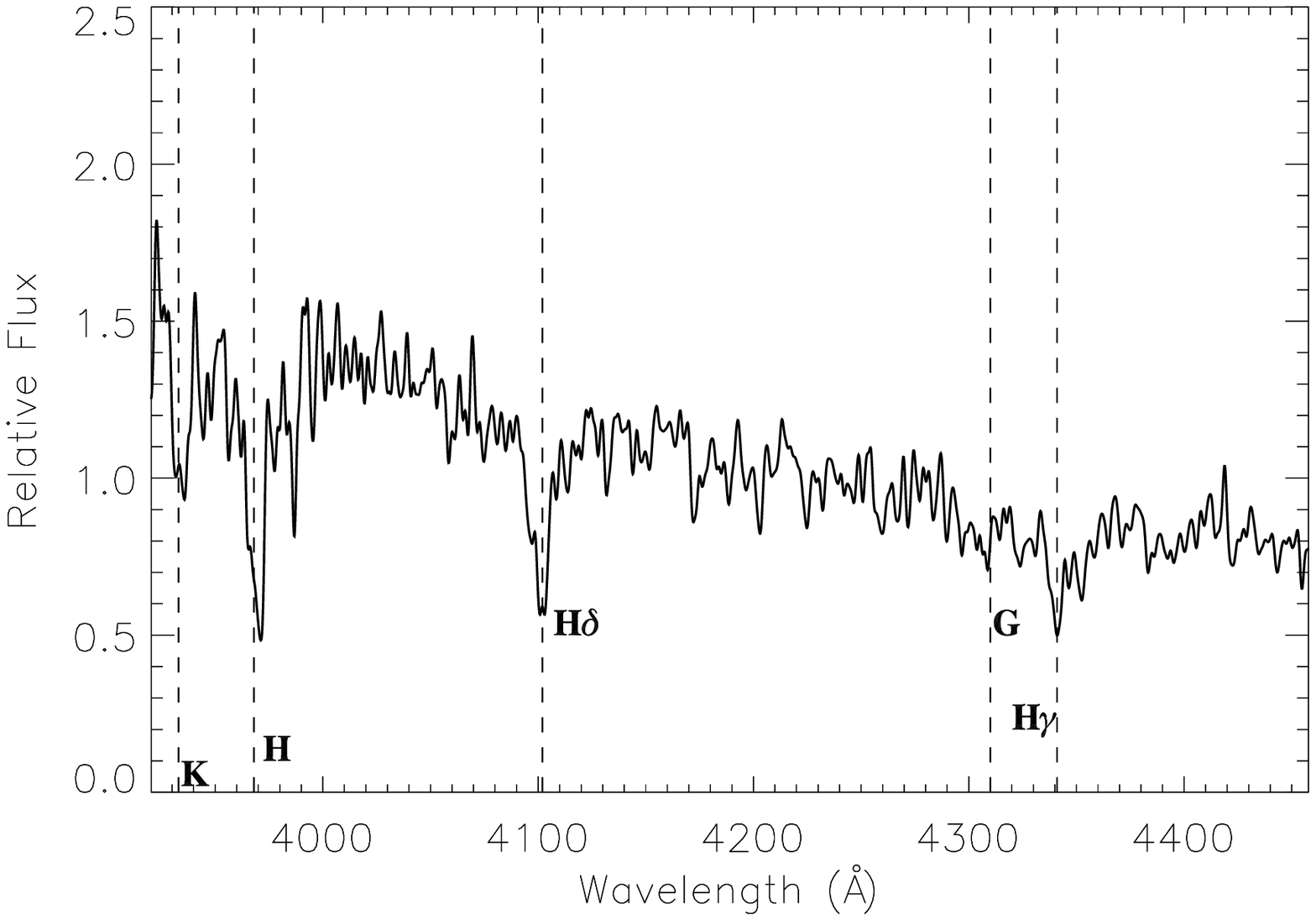}
   \includegraphics[width=0.45\textwidth]{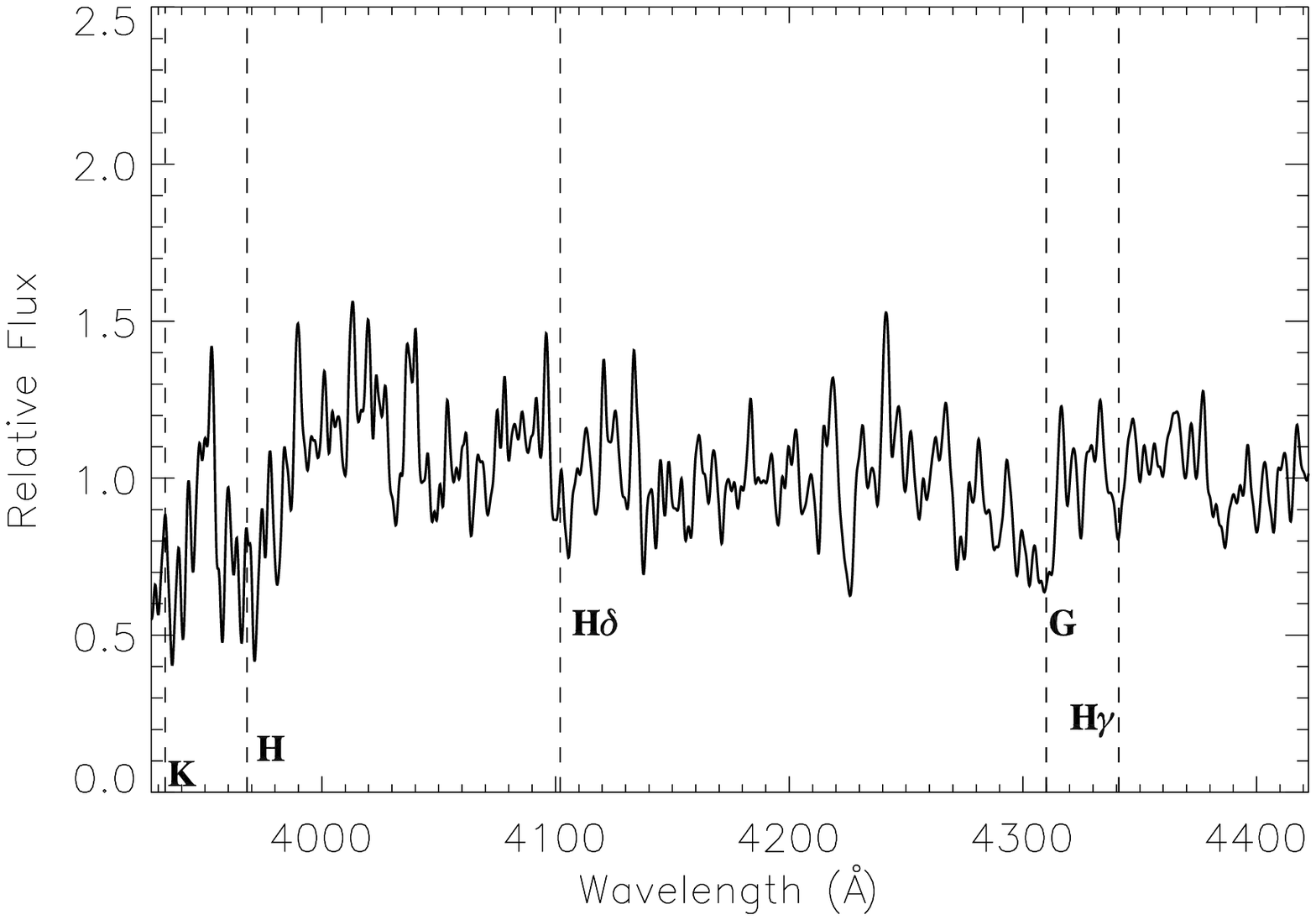}\\
   \includegraphics[width=0.45\textwidth]{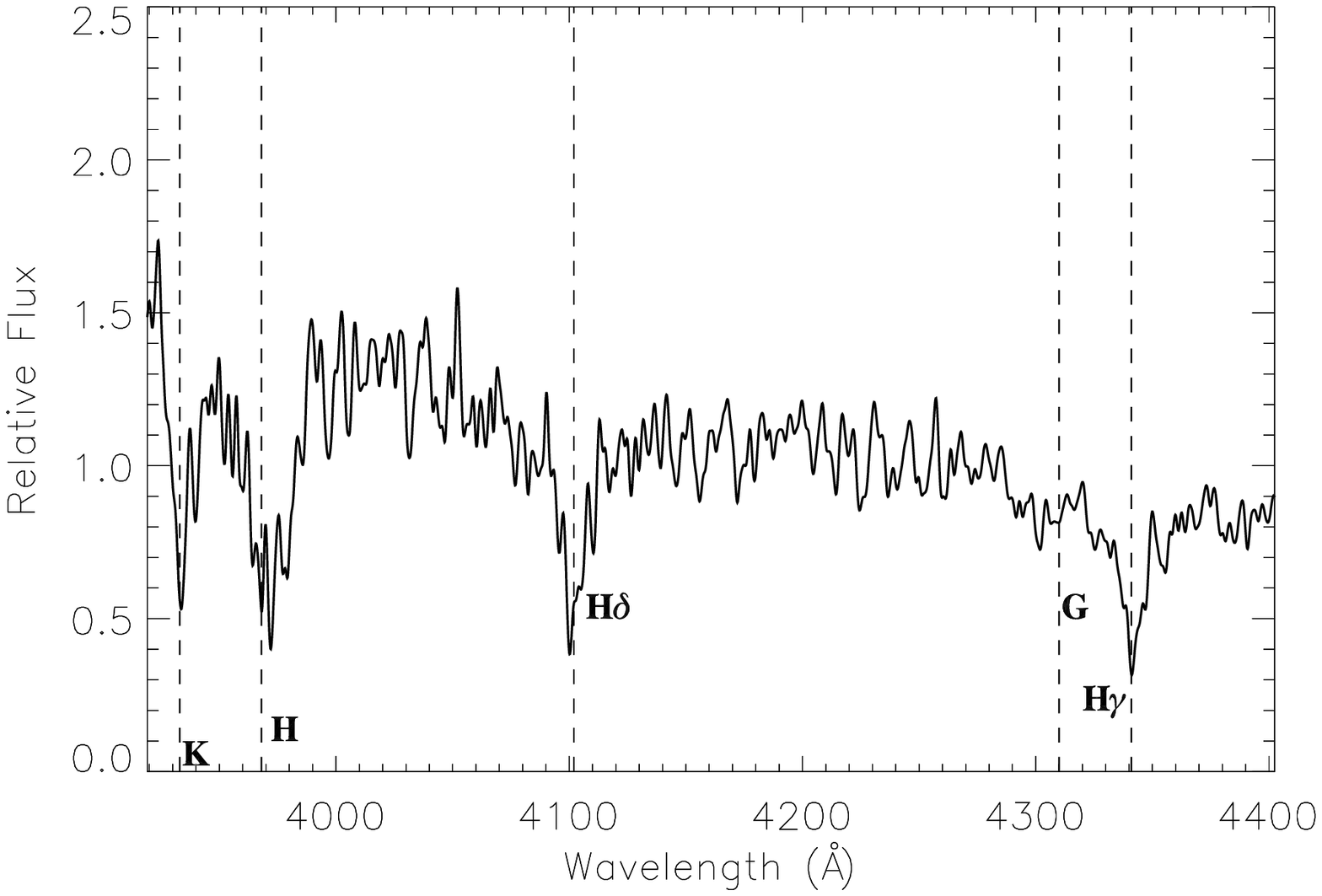}
   \includegraphics[width=0.45\textwidth]{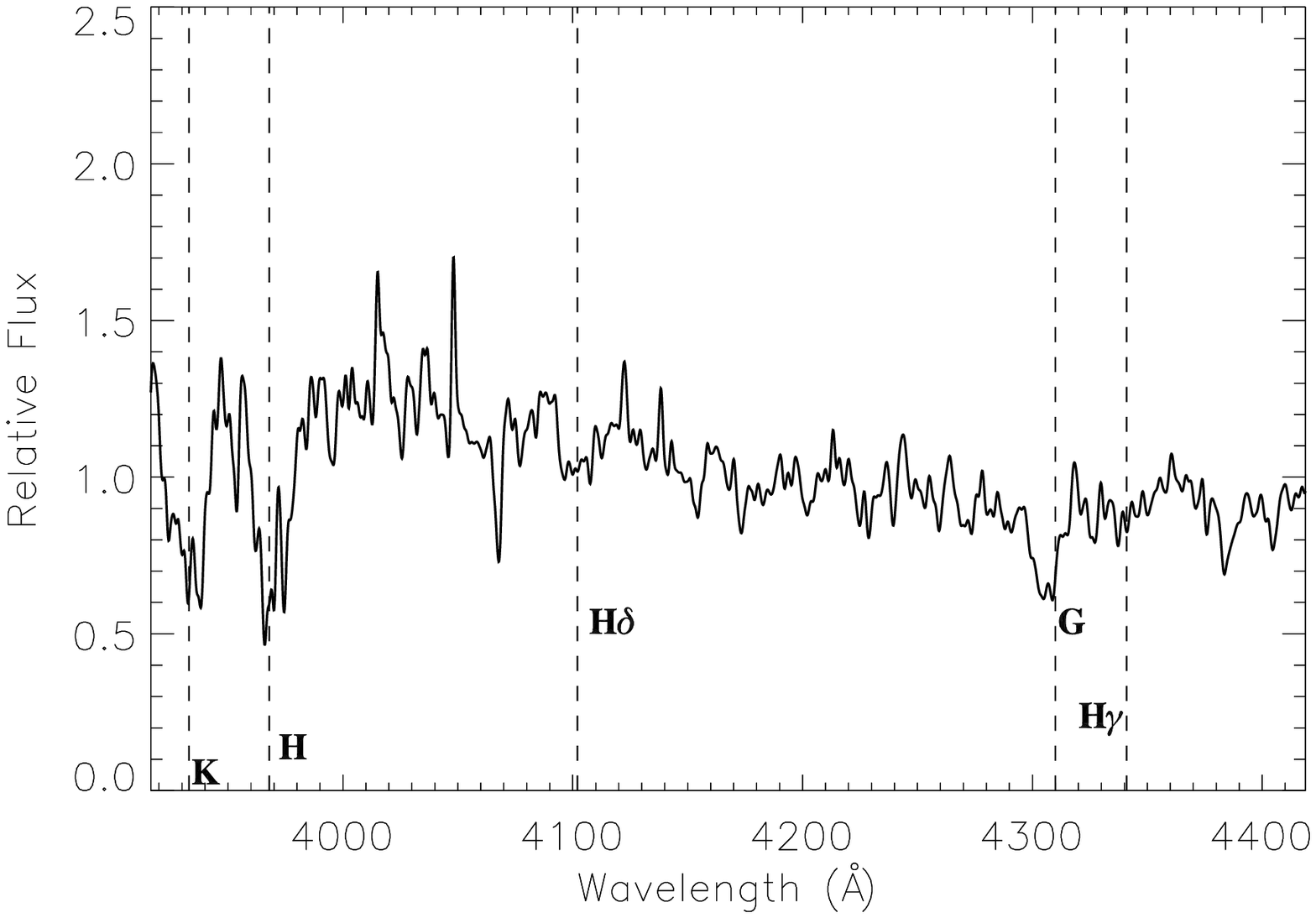}
\end{center}
\vspace{-0.2cm}
\caption{Integrated spectra for a representative sample of our 
galaxies, from left to right and top to bottom: CN191, CN232, CN143 and CN74. 
We provide two examples of targets with prominent Balmer absorption (left) and two targets
without (right). The spectra have been smoothed with a Gaussian 
of FWHM 1\AA{} to improve their presentation. Note that the spectra have not been
flux calibrated.}\label{spectra}
\end{figure*}


\section{ANALYSIS}

We begin by studying the global properties of the sample, by integrating 
the fibres from each IFU to produce a single spectrum per galaxy. For most 
of the sample we combined all the fibres.  However, in a few cases this 
resulted in an excessively-noisy spectrum, and therefore only fibres with 
signal-to-noise ratio (S/N) $\ga 5$\AA$^{-1}$ (defined in selected regions of the 
continuum) were then combined.  
The rejected fibres were always far from the brightest pixel, in the outskirts 
of the target galaxy.

In Figure \ref{spectra} we plot the integrated spectra for a
representative subsample of the galaxies: CN191, CN232, CN143 and CN74. 
The signal-to-noise ratios (S/N) of the integrated spectra were relatively high, reaching
values of $\sim22$\AA$^{-1}$ (CN146).  All of the spectra display the K and H
Calcium lines and the G-band, which are characteristic of an old
population. However, the H$\delta$ and H$\gamma$ absorption lines,
produced by the young, A-star population are only strong in two of the
spectra.  The lack of strong Balmer absorption in the remainder
contrasts with their selection as `k+a' galaxies. Below we measure the
H$\delta$ index of the galaxies in order to quantify the strength of
this feature.

We are also interested in considering spatially-resolved information
from the different regions of the galaxies covered by the IFUs.  In
the majority of the galaxies at least some of the central fibres had
sufficient S/N (reaching values of $\sim 15$\AA$^{-1}$) to be
analysed individually, although the degree to which this is possible
varies between galaxies. For this reason, in addition to performing the
analysis for the individual fibres, in each galaxy we combined all the
pixels immediately adjacent to the brightest one, which we refer to as
the `surroundings' (covering from $\sim$1.6 to 3.2 kpc), and those placed further away, which we define as
the `outskirts' ($\sim$3.2 to 4.8 kpc). In some cases, due to the low S/N in the pixels far
away from the centre, we could not obtain reasonable quality spectra
for the `outskirts'.

To find the centre of each galaxy we built images of the light
distribution in the continuum region between the H$\delta$ feature
and the sky line at $\lambda$5577 for each IFU. The centre
of the galaxy was associated with the brightest (and hence usually
highest S/N) pixel.  Some large and inclined galaxies were
purposefully offset to include their outer regions in the IFU.
However, many of the other galaxies also display offsets from the IFU
centre. These offsets, which are also present in the observations
carried out by P05, are likely a result of inaccuracies in the
astrometry and IFU positioning errors.  They are, however,
significantly smaller than the field of view so do not compromise
the analysis.

\subsection{Indicators of a young population in the `k+a' galaxies}

\subsubsection {Line index measurements}

As explained above, `k+a' spectral features arise from the truncation
of star formation in a galaxy, which may be preceded by a starburst,
and reflect the composite of a young and an old stellar population.
These galaxies are usually identified by the strong Balmer absorption
lines in their spectra. Since the higher-order Balmer lines are less
affected by emission from ionized gas \citep{Osterbrock_1989}, the
most commonly used indicator is the H$\delta$ line at $4102$\AA{},
which is also conveniently located in the optical part of the spectrum at low and
intermediate redshift. Although the selection criteria does vary
depending on the study, `k+a' galaxies are generally selected to have
$\rmn{EW}(\rmn{H}\delta) > 3$\AA{} and no detected emission lines.

The strength of the H$\delta$ absorption line is related to the
mechanism responsible for the `k+a' feature. \citet{Poggianti_1999}
showed that strong H$\delta$ absorption lines ($\rmn{EW}(\rmn{H}\delta) >
4$--$5$\AA{}) can only be caused by the abrupt truncation of star
formation after a starburst.  Lower values of [EW(H$\delta$)] can also
be achieved by a simple truncation of a continuous and regular star
formation in the galaxy.  However, the strength of the H$\delta$ line
subsides with time, so it is difficult to distinguish between recent
truncation and an older one that was preceded by a starburst.

Although we consider more sophisticated indicators of the stellar
population later in this paper, given the importance and simplicity of
the H$\delta$ absorption feature, we first measure the equivalent
width of this line for the sample galaxies. We utilised the redshifts
obtained from the template fits with {\tt pPXF} (Section
\ref{sec:ppxf_analysis}), as listed in Table~\ref{table:2}. Equivalent
widths were measured using the software {\tt INDEXF}
\citep{Cardiel_2010}, which uses the Lick/IDS index definitions of
\citet{Worthey_Ottaviani_1997} to measure the signal within the line
with respect to the neighbouring continuum. To make our results
comparable with those obtained by P05, we use the index H$\delta_{\rm
  F}$, which takes the continuum intervals $4057.25$--$4088.5$\AA{}
and $4114.75$--$4137.25$\AA{} around the central
$4091.00$--$4112.25$\AA{} bandpass.  Errors are estimated from the
propagation of uncertainties in the spectra and the measured radial
velocities.

Four of the galaxies in our sample display emission lines, which would
affect the line index measurement due to the filling of the absorption
lines. To avoid this, for these four galaxies instead of using the
original spectrum we measured the line index on the best fit spectrum
constructed by {\tt pPXF}. This procedure has been shown to produce
very good results by \citet{Johnston_2013a}.

The values of H$\delta_{\rm F}$ for all the galaxies 
in our sample are listed in Table \ref{table:2}. We also list 
the values obtained for the galaxies from the P05 sample.
For galaxies that are present in both samples we obtained 
very similar values, consistent within the given uncertainties.  Hereafter
we used the values measured in our data, because they 
possess higher S/N ratios.

The first surprising finding is the number of galaxies for which we
measure H$\delta_{\rm F}$ lower than $3$\AA{}.  This was already
suggested from the weak Balmer absorption lines apparent in some of
the spectra upon visual inspection, see (Figure \ref{spectra}). These low
values contrast with those expected from the spectroscopic study by
CS87, in which all of our sample showed $\rmn{EW}(\rmn{H}\delta)$ higher than $3$\AA{}.
This discrepancy was also found by P05. It appears that the uncertainties in 
the CS87 H$\delta$ EWs are rather
large, and hence their spectral classifications are only reliable for the
most extreme `k+a' cases.

From our analysis, only seven of the twenty galaxies display $\rmn{EW}(\rmn{H}\delta) > 3$\AA{},
 with three of them also having detected [OII]$\lambda$3727 emission.
If we consider also those with $\rmn{EW}(\rmn{H}\delta)~>~2$~\AA {}, three more galaxies
are included, giving a total of ten. The values obtained in our analysis of the P05 sample are in
reasonable good agreement with what they found, considering 
that each study applied a different method. We only found one 
galaxy, CN849, where we measured a lower value of H$\delta_{\rm F}$ 
(2.3 $\pm$ 0.4) than what they obtained (3.6 $\pm$ 0.3), which in this
case is significant because it changes the galaxy's `k+a' classification.

As mentioned above, four of the galaxies we observed are listed as having
[OII]$\lambda$3727 emission in CS87 and therefore do not meet the
standard `k+a' criteria. Their EW([OII]) values range from 7.6\AA{} to
39.6\AA{}. These are likely reliable emission line identifications.
Three of them are found to have H$\delta_{\rm F} > 3$~\AA{} (CN146,
CN155 and CN243) and they also show signs of emission in H$\gamma$ and H$\delta$ in our
data. However, it is not clear whether these emission lines result
from residual star formation or AGN activity.

\begin{figure}
\begin{center}
   \includegraphics[width=0.45\textwidth]{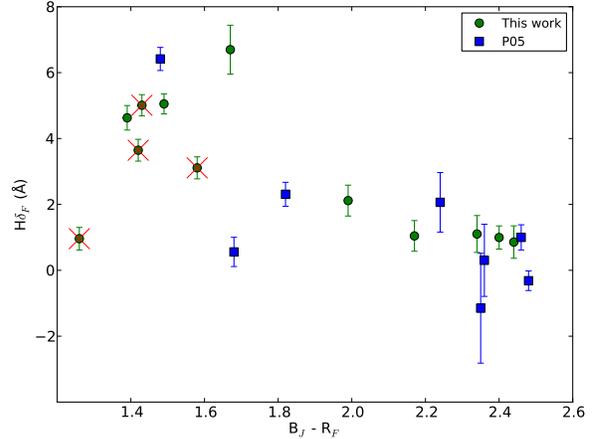}
\end{center}
\vspace{-0.4cm}
\caption{H$\delta_{\rm F}$ versus $B_J-R_F$ colour 
for our entire galaxy sample. In the case of objects that 
were observed by both P05 and ourselves, we only plot our values. Galaxies with detected 
emission in [OII] by CS87 are indicated by a red cross.
}\label{fig:color}
\end{figure}

One would expect that if star formation has been recently truncated in
those galaxies with strong H$\delta$ absorption, they should have
bluer colors due to the presence of the young population. To test
this, in Fig.~\ref{fig:color} we plot H$\delta_{\rm F}$ versus $B_J
- R_F$ for all the sample galaxies.  Objects with strong H$\delta$ absorption
are conspicuously bluer than those with weaker H$\delta$ absorption.
CS87 also present this plot, finding a consistent trend, though
somewhat weaker, presumably due to the larger uncertainties on their
$\rmn{EW}(\rmn{H}\delta)$ estimates.  This trend gives compelling
support that the galaxies in our sample with stronger H$\delta$
absorption, and particularly $\rmn{EW}(\rmn{H}\delta) \ga 2$,
contain younger stellar populations.

In Fig.~\ref{fig:color} we also indicate the galaxies which have observed
[OII] emission.  Recall that, for these galaxies,
$\rmn{EW}(\rmn{H}\delta)$ was measured on the template fits produced
by {\tt pPXF}, rather than the data itself, to avoid the effect of
line-filling.  It is possible that in the case of the bluest galaxy
the line-filling has affected the {\tt pPXF} fit itself, resulting in
an underestimate of $\rmn{EW}(\rmn{H}\delta)$.  These four galaxies
are not strictly `k+a' systems, they may simply be normal star-forming
galaxies, although their high $\rmn{EW}(\rmn{H}\delta)$ might indicate
some recent suppression of their star-formation.  Nevertheless, we retain them
in the analysis because they probably lie just outside the boundaries of the `k+a' class,
 and may provide useful clues on the process by which galaxies become `k+a' systems. 


\subsubsection {A/(AFGKM) and $f_{\rm young}$ measurement}

Although strong H$\delta$ absorption is the standard indicator of a
young population in `k+a' galaxies, this simply reflects the presence
of a substantial stellar population with ages between $0.5$ and $1.5$
Gyr, whose light is dominated by A stars, but an absence of younger
populations containing OB stars, powering nebular
emission from HII regions.  The presence of this intemediate-age stellar population may
also be inferred using other, more quantitative, methods.  One
approach is template fitting, which uses the full wavelength range
available and accounts for the fact that populations of all ages
contribute to $\rmn{EW}(\rmn{H}\delta)$ (and other spectral features).
We use the results of template fits performed using {\tt pPXF}, as
described in Section \ref{sec:ppxf_analysis}.

To estimate the relative proportion of each stellar population, we use
the normalized light-weighted proportions assigned to the various templates in the
best-fitting model.

From the weights obtained using the stellar library templates we
determine the fractions of each stellar type (OBAFGKM) contributing to
the galaxy spectrum. For the fits using the SSP models, we group the
templates into four age bins: `$\rmn{Age} < 0.5$~Gyr', `$0.5 <
\rmn{Age} < 1.5$~Gyr', `$1.5 < \rmn{Age} < 7$~Gyr' and `$\rmn{Age} >
7$~Gyr'.  One expects an approximate correspondence between the
stellar types and SSP ages: stars formed very recently (OB) will
dominate the $\rmn{Age} < 0.5$~Gyr bin, stars with lifetimes
$\sim1$~Gyr (main sequence A and F stars) will dominate the `$0.5 < \rmn{Age} < 1.5$~Gyr' bin,
and longer-lived stars (GKM) will correspond to the two older age
bins.  However, the stellar population templates contain contributions
from stars of all types with lifetimes longer that the SSP age.

To evaluate the fraction of A-type stars we use the ratio A/(AFGKM).
OB stars are excluded from this fraction because their presence is
ill-constrained by our fits, due to their featureless spectra together
with the uncertain flux calibration and limited wavelength range of
our data.  Also, OB stars do not contribute significantly to the
stellar mass of a galaxy.  For the stellar populations, our primary
quantity is the fractional contribution of SSPs with $0.5 < \rmn{Age}
< 1.5$~Gyr over the total, hereafter $f_{\rm
  young}$. The
values of A/(AFGKM) and $f_{\rm young}$, determined with the
integrated spectra for each galaxy, are listed in Table~\ref{table:2}.


\begin{figure}     
\begin{center}
   \includegraphics[width=0.47\textwidth]{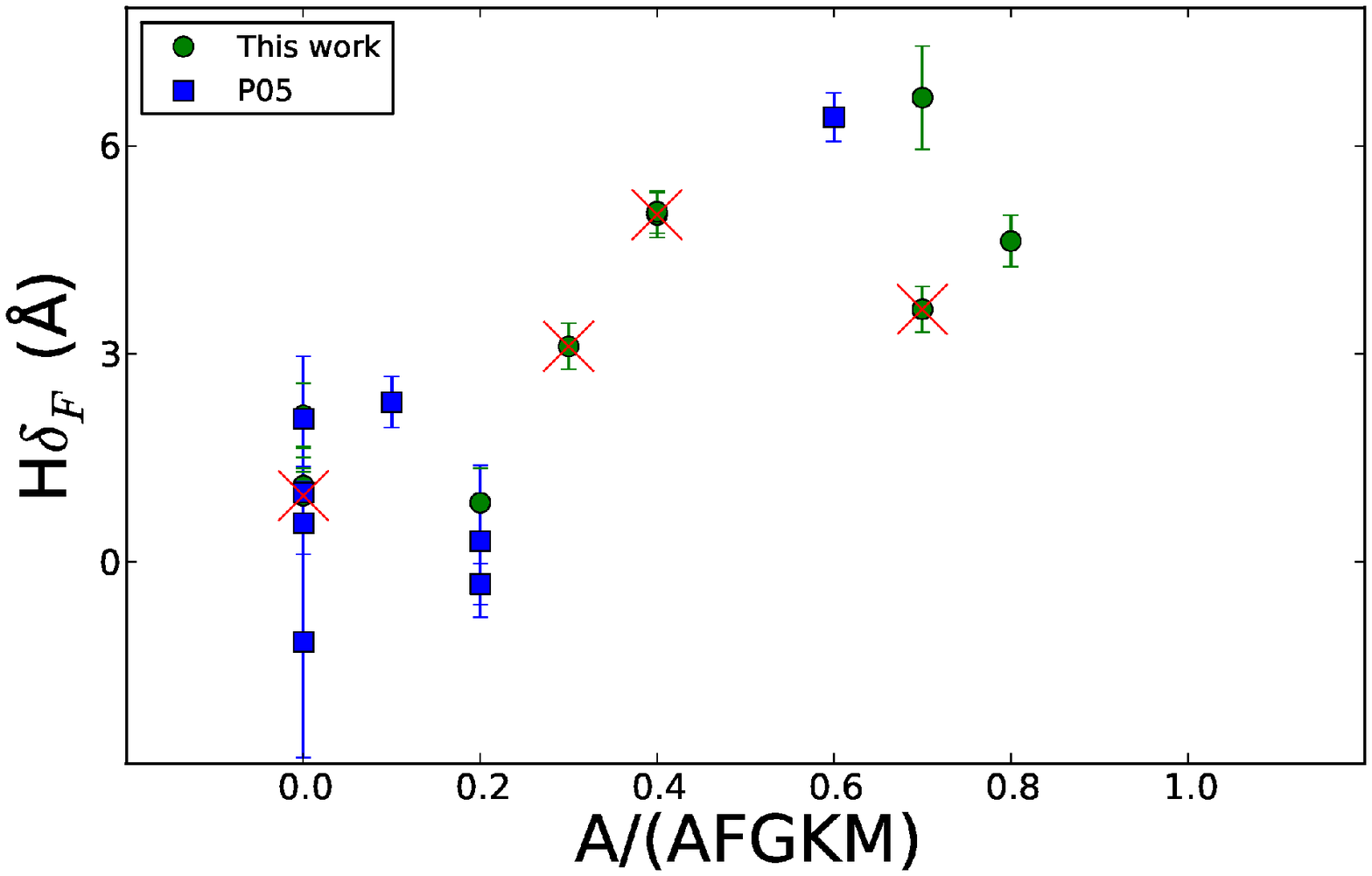}
   \includegraphics[width=0.47\textwidth]{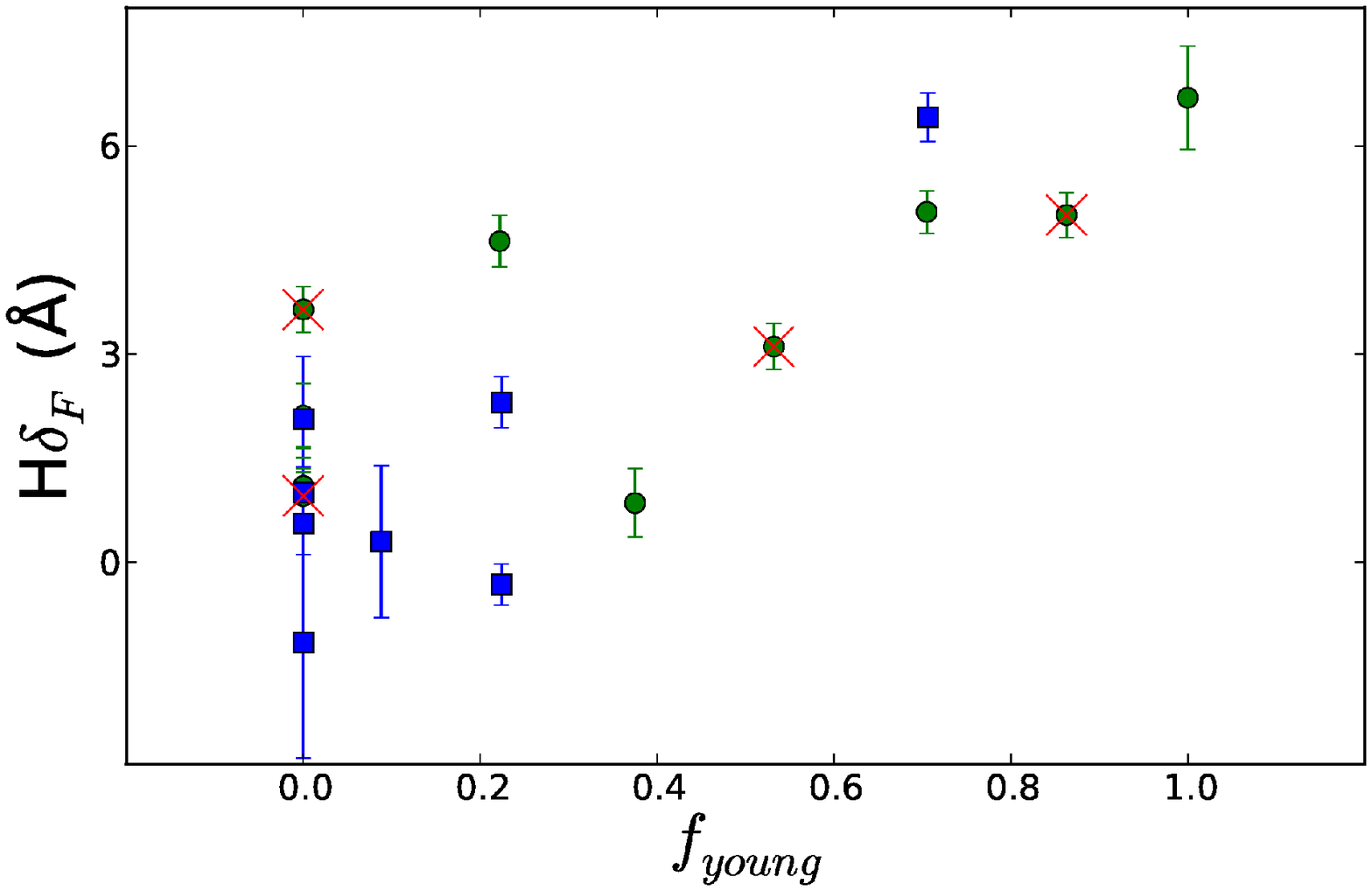}
   \includegraphics[width=0.47\textwidth]{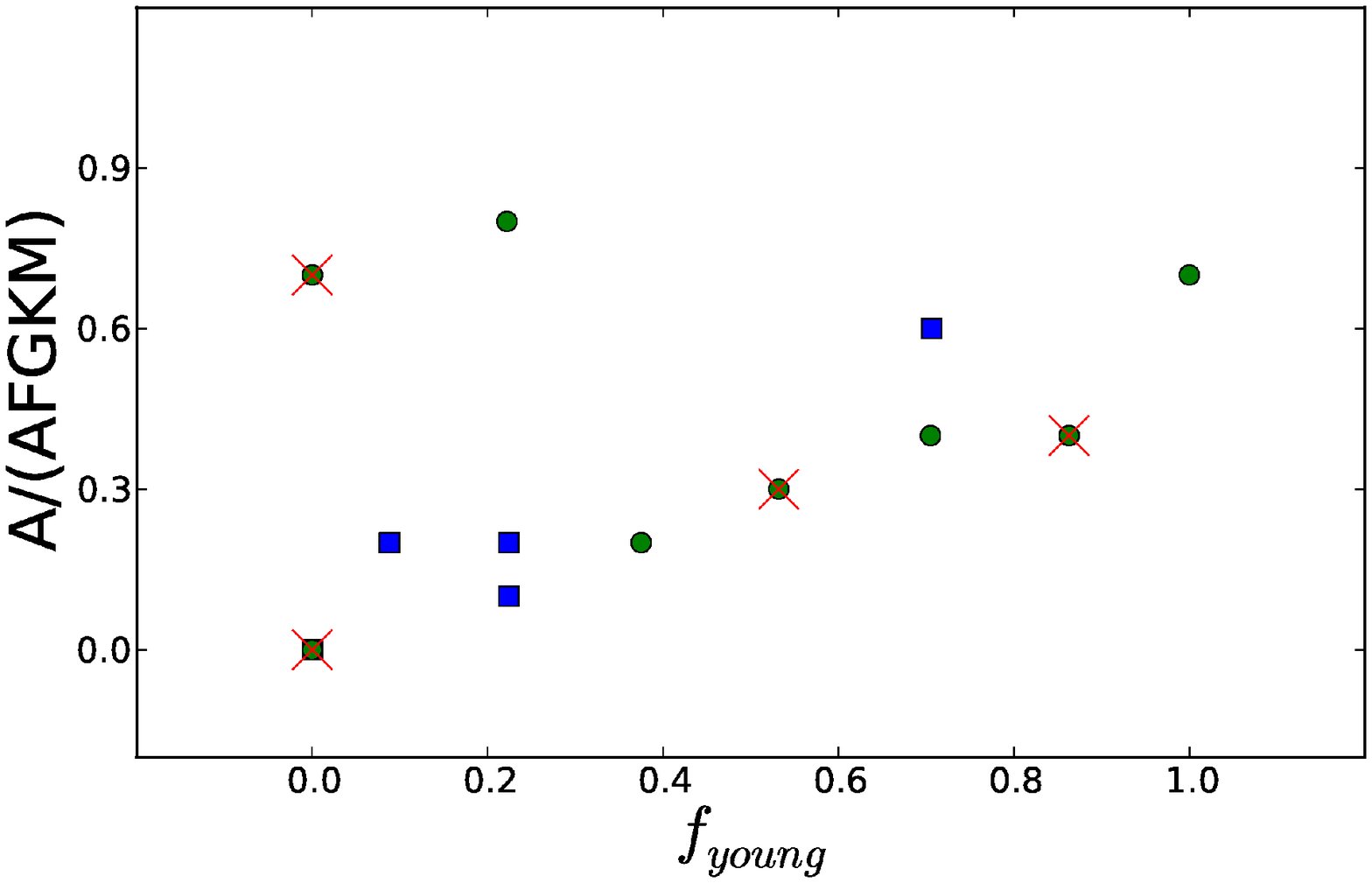}
\end{center}
\vspace{-0.2cm}
\caption{Values of A/(AFGKM), $f_{\rm young}$ and H$\delta_{\rm F}$
  plotted against each other 
for our sample (green circles) and measured by us on spectra obtained by
  P05 (blue squares). For galaxies that are present in both samples we
  only plot values measured in our data because they have higher
  S/N. We only plot error bars in H$\delta_{\rm F}$, as {\tt pPXF}
  does not provide uncertainties on the weights in the best-fitting
  combination of templates. Galaxies with detected emission in [OII]
  by CS87 are indicated by a red cross. In the plot of $f_{\rm young}$ vs A/(AFGKM)
there are fewer visible points because they are superimposed onto each other.}\label{fig:dist}
\end{figure}

Thus now we have three different indicators of the 
presence of a young populations in these galaxies, 
H$\delta_{\rm F}$, A/(AFGKM) and $f_{\rm young}$. 
Comparing these parameters provides a useful indication of their robustness,
 and therefore 
the reliability of using only one of them in cases when
the other ones cannot be obtained. This comparison is done in 
Fig.~\ref{fig:dist}, where we plot A/(AFGKM) and 
$f_{\rm young}$ against H$\delta_{\rm F}$ for our entire 
galaxy sample. Uncertainties on the
H$\delta_{\rm F}$ measurements are indicated, but {\tt pPXF} does not
provide error estimates for the template weights.  We therefore estimate
average uncertainties for A/(AFGKM) and $f_{\rm young}$ 
from the standard deviation of the scatter from a linear correlation 
with respect to H$\delta_{\rm F}$ after subtracting the contribution to their error
 by $\Delta$H$\delta_{\rm F}$. We obtain an uncertainty in both quantities of 0.2.
which is also noted in Table \ref{table:2}.

As expected, there is a good correlation between these
quantities. Galaxies with strong Balmer absorption lines also show
high fractions of A/(AFGKM) and $f_{\rm young}$, while those with weak
H$\delta_{\rm F}$ show very low values of A/(AFGKM) and $f_{\rm
  young}$. The fractions of A/(AFGKM) and $f_{\rm young}$ also present
good correlation between them. More quantitatively, for A/(AFGKM) and H$\delta_{\rm F}$ we
obtain a Spearman's correlation coefficient $\rho = 0.65$, while for
$f_{\rm young}$ and H$\delta_{\rm F}$, $\rho = 0.56$. In the case of 
A/(AFGKM) and $f_{\rm young}$, $\rho = 0.76$. The chance of 
any of these correlations being spurious is $\la 1$ per cent.

\subsection{Spatial distributions}\label{sec:distributions}

As mentioned previously, in addition to providing global information
about a galaxy, integral field spectroscopy allows us to study
properties at smaller spatial scales and hence consider different
regions within a galaxy.  We exploit this possibility by performing the same
analysis described above, but now applied both to the spectra from
individual IFU elements and to combined spectra from the `centre',
`surroundings' and `outskirts' regions of each galaxy.

\begin{figure}     
        \begin{center}
         \includegraphics[width=0.45\textwidth]{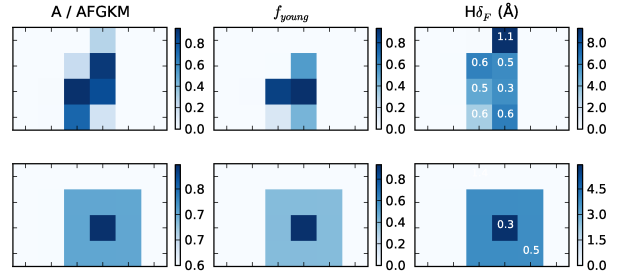}
\end{center}
\vspace{-0.4cm}
\caption{Maps of the individual fibre values of A/(AFGKM), $f_{\rm young}$ and
  H$\delta_{\rm F}$ index (top) and the corresponding values for the integrated regions
`centre' and `surroundings' (bottom) in CN228. Errors of the H$\delta_{\rm F}$
  index are printed over the regions. Each spatial pixel (or spaxel) has a size of 0.52 x 0.52 arcsec$^2$ which
 corresponds to $\sim$ 2.3 x 2.3 kpc$^{2}$ at the redshift of AC114.}
\label{figure:distributions1}
\end{figure}

We have used these results to construct maps of the
three different age indicators, H$\delta_{\rm F}$, A/(AFGKM)
and $f_{\rm young}$, for each galaxy. In many galaxies, due to the S/N being too low in
the `outskirts', only the `centre' and
`surroundings' could be analysed.
Cases where the three integrated regions could be analysed can be found in the Appendix (CN74 and CN849).
An example of this analysis is shown in Fig.~\ref{figure:distributions1}
for the galaxy CN228, where the three indicators show a high concentration of the young population
in the centre of the galaxy. This is particularly clear when considering the `centre`
versus `surroundings` regions.  The values of the individual fibres
for A/(AFGKM) and $f_{\rm young}$ also show a high concentration
towards the centre, while the individual H$\delta_{\rm F}$ are less conclusive.

\begin{figure}    
        \begin{center}
          \includegraphics[width=0.48\textwidth]{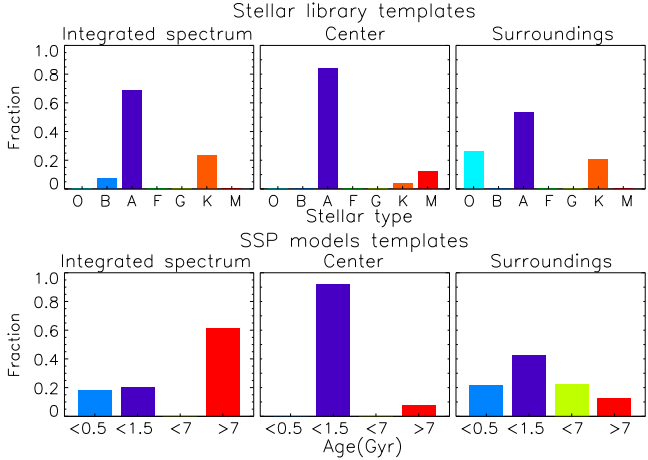}
\end{center}
\vspace{-0.4cm}
\caption{Histograms of stellar type and 
stellar population age obtained with pPXF for the integrated 
spectra, `centre' and `surroundings' of CN228.}\label{figure:distributions2}
\end{figure}

To examine the stellar population in more detail, in
Fig.~\ref{figure:distributions2} we show the normalized distributions
of the different spectral types and SSP ages obtained for
different regions of the same galaxy. The two approaches are broadly
consistent: a prominent fraction of A-stars is associated with a
significant young-age population.

The maps of H$\delta_{\rm F}$, A/(AFGKM) and $f_{\rm young}$ for each
galaxy are our primary source of information regarding the spatial
distributions of the young and old stellar populations.  However, the
maps are difficult to deal with quantitatively, and there is some
subjectivity in identifying the trends they reveal.  We have examined
these maps in detail, and in the Appendix we present qualitative
descriptions of each galaxy, in addition to the maps themselves.

\begin{figure}
\begin{center}
   \includegraphics[width=0.45\textwidth]{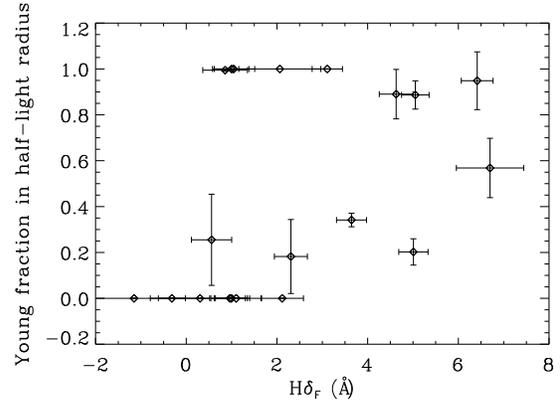}
\end{center}
\vspace{-0.4cm}
\caption{The luminosity-weighted fraction of the young stellar population contained within the half-light radius of the old population plotted against the global H$\delta_{\rm F}$. A value of this fraction larger than $0.5$ indicates that the young population is more concentrated than the old one (see text for details).}\label{fig:scales}
\end{figure}

In an attempt to quantify the differences in the spatial distributions of the young and old stellar populations we
have used these maps to estimate the luminosity-weighted fraction of the young stellar population contained within the half-light radius of the old population. We have assumed exponential intensity profiles for both populations. A value of this fraction larger than $0.5$ indicates that the young population is more concentrated than the old one. Figure~\ref{fig:scales} shows this fraction plotted against the global H$\delta_{\rm F}$ values. There is a large scatter, indicating significant differences in the current properties and formation histories of the galaxies. Nevertheless, galaxies with the strongest H$\delta_{\rm F}$ seem to show some tendency to have more centrally-concentrated young populations. This suggests that the last episode of star formation often took place in the central regions of these galaxies. The sample size, spatial resolution and uncertainties of this study prevent us from reaching a very robust conclusion in this respect, but it is reassuring that our findings are consistent with independent evidence from recent studies of local S0 galaxies \citep{Bedregal_2011, Johnston_2012, Johnston_2013b}.

\begin{table*}
{\scriptsize
\begin{center}
\smallskip
\begin{tabular}{lcccclccl}
\hline
\hline
\noalign{\smallskip}
Target   &   z   &H$\delta_{\rm F}$& A/(AFGKM)     & $f_{young}$   &   $\sigma_{\rm int}$     & V$_{\rm rot}$     & V$_{\rm rot}$/$\sigma_{\rm int}$& Interacting\\
         &       &  (\AA) 	     & ($\pm$ 0.2)	           &	          ($\pm$ 0.2)      & ($\kms$)  & ($\kms$)  & 		 & 	\\
\hline

CN143 	 & 0.310 &  6.7 $\pm$ 0.7    & 0.7 & 1.0   & 107 $\pm$  15 & 133 $\pm$  18 & 1.2 $\pm$ 0.2  & NO	\\
CN191 	 & 0.305 &  5.1 $\pm$ 0.3    & 0.4 & 0.7   &  61 $\pm$   8 & 137 $\pm$  51 & 2.2 $\pm$ 0.9  & NO	\\
CN155$^{*}$& 0.320 &  5.0 $\pm$ 0.3    & 0.4 & 0.9 & 282 $\pm$  72 &  -  	        &  -  		 & YES	\\
CN228 	 & 0.317 &  4.6 $\pm$ 0.4    & 0.8 & 0.2   & 210 $\pm$  66 & 177 $\pm$  38 & 0.8 $\pm$ 0.6  & YES \\
CN146$^{*}$& 0.300 &  3.7 $\pm$ 0.3    & 0.7 & 0.0 & 184 $\pm$  20 &  -  	        &  -  		 & NO   \\
CN243$^{*}$& 0.326 &  3.1 $\pm$ 0.3    & 0.3 & 0.5 & 232 $\pm$  19 &  -  	        &  -  		 & YES	\\
CN254 	 & 0.309 &  2.1 $\pm$ 0.5    & 0.0 & 0.0 & 202 $\pm$  26 & 166 $\pm$  20 & 0.8 $\pm$ 0.2  & NO	\\
CN232 	 & 0.315 &  1.1 $\pm$ 0.6    & 0.0 & 0.0 & 185 $\pm$  32 & 137 $\pm$  55 & 0.7 $\pm$ 0.3  & NO	\\
CN24 	 & 0.322 &  1.0 $\pm$ 0.4    & 0.0 & 0.0 &  93 $\pm$  12 &  -  	        & -  		 & NO	\\
CN74$^{*}$ & 0.316 &  1.0 $\pm$ 0.3    & 0.0 & 0.0 & 239 $\pm$  88 & 152 $\pm$ 75 & 0.6 $\pm$ 1.6  & NO	\\
CN187 	 & 0.308 &  1.0 $\pm$ 0.5    & 0.0 & 0.0 & 202 $\pm$  24 &  -  	        &  -  		 & NO	\\
CN119 	 & 0.308 &  0.9 $\pm$ 0.5    & 0.2 & 0.4 & 144 $\pm$ 134 &  -  	        & - 		 & NO	\\
\hline
CN22 	 & 0.336 &  6.4 $\pm$ 0.4    & 0.6 & 0.7 &  83 $\pm$  22 & 83 $\pm$  21 & 1.0 $\pm$ 1.0  & YES	\\
CN849 	 & 0.324 &  2.3 $\pm$ 0.4    & 0.1 & 0.2 & 189 $\pm$   7 & 86 $\pm$  11 & 0.5 $\pm$ 0.3  & YES	\\
CN89 	 & 0.317 &  2.1 $\pm$ 0.9    & 0.0 & 0.0 &  94 $\pm$  40 &  -  	        &  -  		 & NO	\\
CN247 	 & 0.319 &  1.0 $\pm$ 0.4    & 0.0 & 0.0 & 214 $\pm$  10 &  -  	        &  - 		 & YES	\\
CN667 	 & 0.312 &  0.6 $\pm$ 0.4    & 0.0 & 0.0 & 128 $\pm$  18 &  -  	        &  -  		 & YES	\\
CN229 	 & 0.320 &  0.3 $\pm$ 1.1    & 0.2 & 0.1 & 186 $\pm$  97 &  -  	        &  -  		 & NO	\\
CN4 	 & 0.308 & -0.3 $\pm$ 0.3    & 0.2 & 0.2 & 185 $\pm$  20 &  -  	        &  -  		 & NO	\\
CN858 	 & 0.312 & -1.2 $\pm$ 1.7    & 0.0 & 0.0 & 182 $\pm$  52 &  -  	        & - 		 & NO	\\
\hline\hline                                                                                                              
\end{tabular} 
\caption{Galaxy ID, redshifts, young population indicators, velocity dispersions, V$_{\rm rot}$/$\sigma$ and state of interaction for the galaxies in
our sample (top) and the P05 sample (bottom). Note that morphology and colour are listed in Table \ref{table:1}. Galaxies labeled with * have [OII]$\lambda$3727 detected emission by CS87.}\label{table:2} 
\end{center}
}
\end{table*}


\subsection{Kinematics}
\label{sec:kinematics}

If spiral galaxies are being transformed into S0s by any of the
processes discussed in the introduction, in addition to the changes in
stellar populations considered above, their kinematics may also be
affected.  The kinematics of the `k+a' galaxies in our sample can
therefore indicate what mechanisms are responsible for the truncation
of their star formation.  If the process acts primarily to starve a
spiral galaxy of its gas supply, the disk rotation should be preserved
in the resulting galaxy. However, if a merger is involved, the remnant
would be expected to show more random motions.

The kinematics of the galaxies analysed here were extracted using the
software {\tt pPXF}, as explained in Section \ref{sec:ppxf_analysis}.
First of all, we obtained a value of the overall velocity dispersion
$\sigma_{\rm int}$ for the integrated spectrum of each galaxy, which are listed
in \ref{table:2}.  It should be borne in mind that for the galaxies
with detected emission by CS87, the measurement of $\sigma_{\rm int}$ might be
affected by the filling of the absorption lines due to emission.  A
wide range of values of $\sigma_{\rm int}$ are found, from $\sim 60$ to $\sim
280 \,\kms$.  These overall $\sigma_{\rm int}$ include contributions from both
rotational and random motions, which we will attempt to separate below.

\begin{figure}
\begin{center}
   \includegraphics[width=0.45\textwidth]{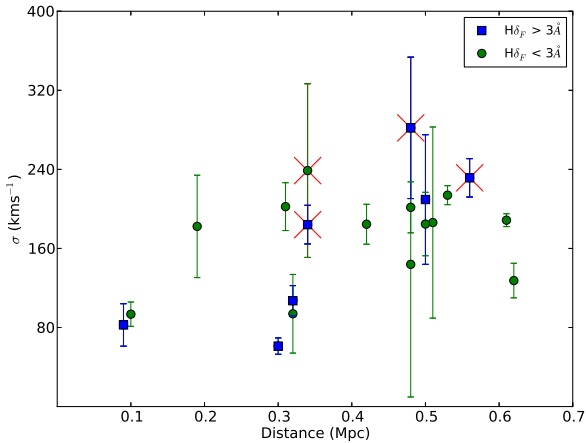}
\end{center}
\vspace{-0.4cm}
\caption{Velocity dispersion $\sigma$ vs projected distance to the centre of the 
cluster for galaxies with H$\delta_{\rm F}$ $>$ 3\AA{ }(blue squares) and 
H$\delta_{\rm F}$ $<$ 3\AA{ } (green circles). For those 
galaxies that are observed by both P05 and ourselves we plot the 
mean value. Galaxies with detected 
emission in [OII] by CS87 have a red cross overplotted.}\label{figure:distance}
\end{figure}

If the gas and the kinematics of the galaxies are being affected by
the cluster environment, one would expect galaxies closer to the
cluster centre to show different behaviour to those that are further
out, as found by \citet{Yara_2011}. To test this, we consider $\sigma_{\rm int}$
as a function of the projected distance from the cluster centre (Table
\ref{table:1}), which is plotted in Figure \ref{figure:distance}. For
the full sample there does not appear to be any clear trend. However,
if we separate galaxies with high ($\ge$ 3\AA{}) and low ($<$ 3\AA{}) 
H$\delta_{\rm F}$, we see that those with high H$\delta_{\rm F}$ (blue
squares) present a strong trend. High H$\delta_{\rm F}$ galaxies have
higher $\sigma_{\rm int}$ the further they are from the cluster centre, while
those with low H$\delta_{\rm F}$ (green circles) show little change
with position.  Some of the high-H$\delta_{\rm F}$ and high-$\sigma_{\rm int}$
galaxies display emission-lines (red crosses), which may make the
estimation of $\sigma_{\rm int}$ unreliable.  However, if we remove them from
the plot, we see that the trend remains.

\begin{figure}
        \begin{center}
         \includegraphics[width=0.45\textwidth]{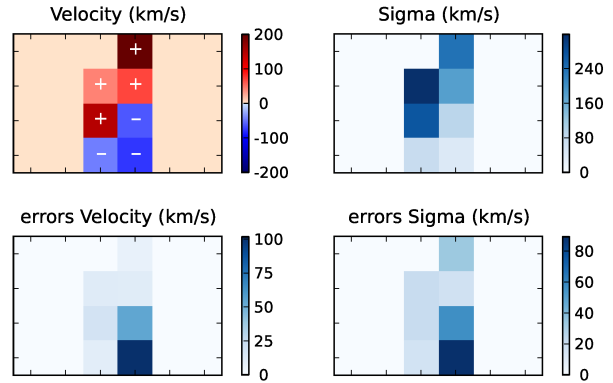}
\end{center}
\vspace{-0.4cm}
\caption{Example of the radial velocity and velocity 
dispersion maps, with errors plotted below. The plus (`+') and minus (`-') 
symbols indicate 
the direction of rotation. In this image
 we show the example of CN228, showing a clear pattern 
of rotation and with similar
values of $\sigma$ along the galaxy.} \label{velocities}
\end{figure}

We now turn our attention to the kinematics of the galaxies on smaller
scales, which can be studied using the outputs of fits performed to
the individual IFU fibres. We construct line-of-sight velocity, $V_{\rm obs}$, and
velocity dispersion, $\sigma$, maps of the galaxies, in a similar manner to those
for the young population indicators. An example is shown in Figure
\ref{velocities} for the galaxy CN228. 

We have studied these maps for signs of rotation and differences in
velocity dispersion between the central and surrounding pixels. One
problem we had to face here was that the `good' fibres were not always
distributed around the brightest pixel in the IFU, and it was sometimes
difficult to identify patterns of rotation or velocity dispersion. 
Since the observed velocity is $V_{\rm obs}~=~V_{\rm rot}{\rm sin}i$,
 where $V_{\rm rot}$ is the rotational velocity and $i$ the inclination of the galaxy,
we need to know the galaxy inclination in order to obtain the actual rotational
velocity. The inclination was therefore determined by the apparent ellipticity 
obtained by fitting an ellipse to the HST/WFPC2 images using the
{\tt IRAF} task {\tt ELLIPSE}. In the case presented in Fig.~\ref{velocities}, we can see a clear
pattern typical of rotation, with $V_{\rm rot} = 177 \pm  38\,\kms$.
The distribution of $\sigma$ is roughly flat.

Previous studies of the kinematics of `k+a' galaxies have found
significant rotation in many of them
\citep{Franx_1993,Caldwell_1996,Pracy_2009,Swinbank_2011, Pracy_2013}, although
some are found to be mainly pressure-supported \citep{Norton_2001}. We
attempted our kinematic analysis in all the galaxies, including the
observations of P05, and found that at least 8 galaxies display
rotation, with values of $V_{\rm rot} \sim 85$--$180 \,\kms$.

The measured values of $\sigma_{\rm int}$ and $V_{\rm rot} $ are listed in Table
\ref{table:2}, along with their ratio ($V_{\rm rot} / \sigma_{\rm int}$), which
indicates whether a galaxy is a rotationally ($>1$) or pressure ($<1$)
supported system. Using this last parameter, we see that 2 of the
systems displaying rotation are clearly rotationally supported,
typical of disk-like systems, while 5 show $V_{\rm rot} / \sigma_{\rm int} < 1$
indicating they are dominated by random motions. Coming back to Figure \ref{figure:distance}, now we are able to
establish if the high values of $\sigma_{\rm int}$ found for some galaxies are
due to rotation or to random motions. From the 10 galaxies with
H$\delta_{\rm F} \ge 2$\AA{}, rotation is detected in 6 of them and
dominant in 2 of these. However, the amount of rotation in galaxies
far from the centre, in particular CN254 and CN228 is
conspicuously higher ($\ge 160 \,\kms$) than in those closer to the
centre such as CN143 and CN191 ($< 140 \,\kms$). The observed trend to
lower internal velocities with decreasing distance from the cluster
core may therefore indicate a trend to less regular kinematics, and hence
environmentally induced disturbances in the centre of the cluster.

\subsection{Kinematic decomposition}

The kinematics studied in the previous section are derived assuming
that all stellar populations contributing to a spectrum have the same
kinematics.  However, our data affords the possibility of measuring
the kinematics of the young and old populations in `k+a' galaxies
separately (e.g., \citealt{Franx_1993, Norton_2001}).  Separated
kinematics offer a further method of distinguishing between the
mechanisms responsible for `k+a' signature.  Rotation in the young
components implies it is in a disk and that the galaxy has not been
subject to a violent process, particularly if the young population
kinematics and distribution are consistent with the older population.
On the other hand, a pressure-supported young population implies that
a significant interaction has occurred.  The degree of rotational
support in the old population may then indicate the strength of this
interaction.

In order to study the kinematics of the two different populations, we
modified the {\tt pPXF} algorithm in such a way that it could fit two
different stellar templates to one spectrum simultaneously, convolving
each one with different radial velocities and velocity dispersions.
The same modified algorithm has been used to study a galaxy with two
counter-rotating disks by \citet{Johnston_2013a}, with good results. In
our case, we used a set of templates containing A-stars and K-stars
with different metallicities so that {\tt pPXF} could clearly distinguish
between the two populations.

Decomposing the kinematics is very challenging, and requires higher
signal-to-noise than available in most of the individual IFU elements.
The decomposition was therefore attempted on coadded spectra
corresponding to three regions for each galaxy, the centre and both
sides, where the orientation of each galaxy is judged from the
kinematic maps from Section~\ref{sec:kinematics}.  We found that the
algorithm was sometimes sensitive to the initial values of $V$ and
$\sigma$ used.  We therefore varied these input values and, in order
to be considered robust, the outputs of the fits were required to
remain constant for a wide range of initial values.

\begin{figure}
        \begin{center}
          \includegraphics[width=0.45\textwidth]{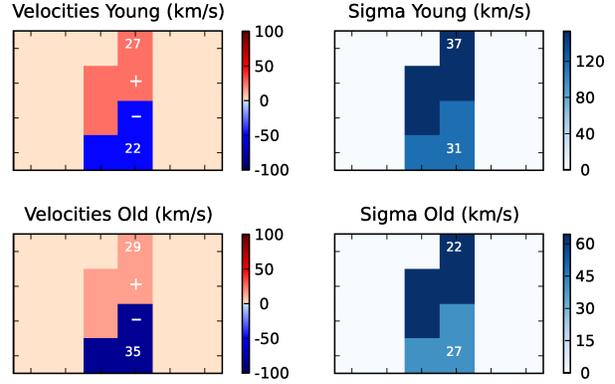}
\end{center}
\vspace{-0.4cm}
\caption{Kinematic decomposition of the young and old stellar
  populations in CN228, obtained using our two-component 
fitting method. Mean velocity and velocity dispersion values are 
presented in the IFU image. The plus (`+') and minus (`-') 
symbols indicate 
the direction of rotation. Errors in the fits are plotted over 
the corresponding regions.}\label{figure:decomposition}
\end{figure}

The results are presented in terms of $V$ and $\sigma$ maps in a
similar manner to the previous section.  As an example, the kinematic
decomposition of CN228 is shown in Figure \ref{figure:decomposition}.
In this case, the galaxy is a composite of two populations with
similar patterns of rotation, while the young population displays
higher values of $\sigma$ than the old population, throughout the
galaxy. As shown previously in Figure \ref{figure:distributions1},
CN228 shows central concentration of the young population in the distribution of the three
indicators H$\delta_{\rm F}$, A/(AFGKM) and $f_{young}$, impliying a
concentration of the young population in the centre of the
galaxy. Now, adding the information provided by the kinematic
decomposition, the fact that this galaxy and others show similar
rotation between the young and old population seems to indicate that
these were fairly normal disk galaxies which have not experienced a
major merger or dominant central starburst. However, the higher
$\sigma$ suggests that they have experienced an interaction which
increased the random motions in the gas from which the last population
of stars was formed.

In total, three disk galaxies with H$\delta_{\rm F}$ $\ge$ 3\AA{} 
(CN228, CN146 and CN191) and two with H$\delta_{\rm F}$ $\ge$ 2\AA{} 
(CN254, CN849) could be kinematically decomposed into two populations. 
In four of these cases both the young and old populations were found to have 
similar patterns of rotation, whereas no clear pattern 
was found in the remaining one (CN146). 


The $\sigma$ values obtained for the two populations display a variety
of behaviours, both in terms of their relative strength and their
radial gradients. In the case of CN228 discussed above, the $\sigma$ of 
the young stars is higher than that of the old population, which 
suggests that this disky `k+a' galaxy may have experienced a recent
interaction, which has increased the random motions of their cold gas,
from which the latest generation of stars have formed, but had less
effect on their previously existing stellar populations. Thus, the process 
cannot be purely gravitational, since the old population is not perturbed, 
and it must be affecting only the gas \citep{Yara_2011}. A more detailed analysis 
of the kinematic decomposition for each galaxy 
is presented in the Appendix.


\subsection {Influence of interactions}

Dynamically interacting galaxies are often observed to be experiencing
a starburst (e.g. \citealt{Keel_1985}).  Simulations have long
suggested that mergers and interactions can cause gas in a galaxy disk
to lose angular momentum and fall toward the centre of the galaxy,
potentially fueling a central starburst
\citep{Barnes_1991,Mihos_1996,Bekki_2005}. However, observations often
find that interactions promote star formation throughout the galaxies
involved (e.g. \citealt{Kennicutt_1987,Elmegreen_2006}), not just in
the nuclear region.  This can now be reproduced by models which pay
closer attention to the role of shock-induced star-formation
(e.g. \citealt{Chien_2010,Teyssier_2010}).

Assuming that the starburst process occurs faster than the
replenishment of the gas disk via infall, or alternatively that such
infall is suppressed, then following the starburst the galaxy will
cease star formation.  The resulting galaxy will therefore display a
k+a spectrum for a time.

The importance of mergers and interactions as the origin of the `k+a'
feature is supported by studies which find that `k+a' galaxies (of all
morphologies) are more likely to be found with a companion galaxy,
when compared to normal galaxies \citep{Goto_2003, Goto_2005,
  Yamauchi_2008, Pracy_2012}.  For example, in their catalogue of
k+a and their companion galaxies, \citet{Yamauchi_2008} found that
k+a galaxies were 54 per cent more likely than normal galaxies to
have a significant companion.  Similarly, the two `k+a' galaxies with
late-type morphology and with a central concentration of the young
population studied by \citet{Pracy_2012} have nearby companions and
could be experiencing tidal interactions.  However, note that all of
these results are based on the general `k+a' population, and thus may
differ from the disky, cluster `k+a' population considered in this
paper.  We have therefore looked for evidence of interactions in the
sample.

In Table \ref{table:2} we have included a column specifying whether each
galaxy displays indications of interacting with other objects.  This was
evaluated by visual inspection of the HST/WFPC2 images of the AC114
cluster. Of the twenty galaxies in our sample, seven have a close companion
and show clear signs of a merger or interaction. The remainder appear fairly
isolated and undistorted. However, the fact that a galaxy does not appear to be
currently interacting does not rule out such a process as the cause of
a `k+a' feature. The spectral `k+a' signature can last for up to $1.5$
Gyr, which is enough time for an interacting galaxy to have moved to a
completely different region of the cluster and any distortion feature
might have faded. 

To test if interactions have any influence in the properties of the
galaxies, we looked for any kind of correlation with any of the
results obtained so far in this study. Of the ten galaxies with
H$\delta$ $>$ 2\AA, five show signs of interaction.  Of the seven
H$\delta$ $>$ 2\AA{} galaxies with disky morphology and usable spatial
information, three have centrally concentrated young populations
(CN155, CN228 and 849) and all of these show evidence for
interactions.  In contrast, the four disk galaxies with their
young population extended throughout the galaxy do not show any sign
of mergers or interactions.

This finding strongly supports a link between dynamical
interactions and a centrally concentrated starburst in disky, cluster 
k+a galaxies.  The remainder, with an apparently less
concentrated young stellar population may simply be the result of
weaker or older interactions, or caused by an alternative mechanism.
However, the strength of H$\delta_{\rm F}$ for the interacting
and non-interacting galaxies does not differ significantly.
 
\section{Discussion}\label{section:discussion}

Our analysis reveals that disky `k+a' galaxies in
intermediate-redshift clusters are a mixed population.  However,
despite the small sample size, we do see some consistent behaviour in a
number of important respects. These results are robust to changes in
the way we quantify the presence and kinematics of the young and old
stellar populations.

The young stellar populations within our sample galaxies are always
either distributed similarly to, or more compactly than, the older
population.  Importantly, however, they are rarely consistent with
being purely confined to the galaxy nucleus.  Furthermore, the young
stars often display rotational kinematics corresponding to the rest of
the galaxy, implying they are located in the disk.  However, there are
some indications that their velocity dispersions are somewhat greater
than in normal spiral galaxies.

Together these results suggest that the young stellar component formed
in an extended disk, in a manner similar to previous generations of
stars in these galaxies.  It is not associated with the aftermath of a
nuclear starburst, nor star formation in tidally accreted material.
However the gas from which the latest stars formed was typically more
centrally concentrated than that from which their predecessors were born.

The scenario presented by our data can be brought together with many
other pieces of observational evidence to support a consistent picture
describing the evolution of the majority of disk galaxies in
intermediate-redshift clusters and groups.

Firstly, we note that any satellite galaxy within a larger halo,
particularly one massive enough to have developed a quasi-static hot
atmosphere \citep{Rees_1977}, is very likely to have its own gas halo rapidly removed
by interactions with the host halo's intergalactic medium and tidal
field, via the mechanisms discussed in the introduction.  The
environmental removal
of HI gas reservoirs is observed both locally (e.g., \citealt{Vogt_2004b}) and
at intermediate redshift (e.g., \citealt{Yara_2012}).  Star-forming
galaxies entering a dense environment (i.e. becoming satellites: low mass galaxies in
groups and higher mass galaxies in clusters), would therefore be
expected to gradually decrease their star-formation rate as they consume
their remaining supply of dense gas.

However, a gradual decline in the star formation rates of star-forming
galaxies in dense environments is at odds with results from large
surveys.  The colours and H$\alpha$ equivalent widths of star-forming
galaxies are invariant with environment (e.g.,
\citealt{Balogh_2004,Balogh_2004_colour,Baldry_2006,Bamford_2008}),
although the relative proportions of blue versus red or star-forming
versus passive galaxies vary substantially.  This strongly implies
that galaxies must rapidly transform from star-forming to passive,
such that a transition population is not seen. The
transformation mechanism cannot be particularly violent, as many
galaxies become passive whilst maintaining their disk morphology, first as
red spirals, and then as lenticulars (e.g.,
\citealt{Lane_2007,Bamford_2009,Maltby_2012}).  We must therefore reconcile the
need for a rapid transformation in terms of observed colour and
emission-line properties, with the requirement that the mechanism only
act relatively gently on galaxy structure.

Star-forming galaxies are observed in environments of all densities,
though they become much rarer in dense regions.  However, it is not
yet clear whether those star-forming galaxies which appear to inhabit
dense regions are simply the result of projection effects, or whether
some galaxies are able to maintain their star formation, at least for
a while, in such extreme environments.  The former would imply that
the transition from star-forming to passive is driven by a
deterministic mechanism, specific to particular environments, whereas
the latter could would permit something more stochastic in nature, in
which the effect of environment is simply to increase the likelihood
of such a transition \citep{Peng_2010}.

A stochastic mechanism, which is not directly related to a galaxy's
broad-scale environment, is supported by the observation that the proportions of
red or passive galaxies show trends across a wide range of
environmental density, and that galaxies with truncated star-formation
are often associated with groups
\citep{Moran_2007,Poggianti_2009a,Wilman_2009,Lackner_2013}, which also host
normal star-forming galaxies.

The reality is probably a combination of the deterministic and
stochastic pictures, for example a mechanism whose effectiveness depends
sensitively on the detailed small-scale substructure of the
environment and a galaxy's orbit through it (e.g.,
\citealt{Font_2008,Peng_2012}).  In any case, the deterministic removal
of a galaxy's gas halo soon after it becomes a satellite makes the
galaxy more vulnerable, helping to reduce the timescale of any
tranformation instigated by a subsequent mechanism.

An initial enhancement of star-formation efficiency early in the
star-forming-to-passive transformation process will effectively reduce
the observability of the transition.  The increased star formation
efficiency would balance the effect of the declining fuel supply,
maintaining the appearance of normality, until the fuel supply is
entirely depleted.  The galaxy would then immediately cease
star-formation and rapidly appear passive.

Briefly enhanced or extended star-formation in the central regions of
cluster spirals is supported by our results, as well as the prevelance of
cluster galaxies with `k+a' spectral types \citep{Poggianti_2009a} and
more centrally concentrated young populations in spirals
\citep{Caldwell_1996,Koopmann_2004a,Koopmann_2004b,Vogt_2004b,Vogt_2004c,Crowl_2008,Rose_2010,Bamford_2007,Yara_2011,Boesch_2013}
and S0s (\citealt{Bedregal_2011,Johnston_2012}; contrary to earlier
results, e.g., \citealt{Fisher_1996}), as well as hints of a
brightened population in the Tully-Fisher relation
\citep{Bamford_2005,Boesch_2013b}.  Simulations demonstrate similar
behaviour (e.g., \citealt{Kronberger_2008}). The process responsible 
of a more centrally concentrated young population could be either 
fading of the external parts of the galaxies or pushing the gas inwards.

Recent studies have found disturbed kinematics in the emission-line
gas in cluster spirals, from which their final generation of stars
would be expected to form \citep{Yara_2011,Boesch_2013}.  The
increased central concentration of the young population in many of our
galaxies is certainly consistent with a decrease in the degree of
rotational support. Unfortunately, the quality of our data make it hard to
directly determine whether the relative velocity dispersion of the
young stars in the cluster spirals is higher than that of the old
stellar populations.  However, together these results suggest that future studies of
cluster S0s may expect to find that the most recent disk stellar
population has a smaller scalelength (and possibly greater
scaleheight) compared to previous generations, implying the presence of a
young, small, thick disk.  Such a feature may also be interpreted as a lense or
additional exponential bulge. 

Dust may also play a role in accelerating the progression of the
observational signatures that would be associated with a transition.
The central concentration of star-formation, as described above, to
the dustier inner regions of galaxies \citep{Driver_2007} results in a
greater fraction of that star formation being obscured from optical
indicators \citep{Wolf_2009}.  The transition stage may thus be hidden from
optical studies, but a population of dusty, red galaxies forming stars
at a significant, though possibly suppressed, rate is revealed by observations at longer wavelengths
\citep{Gallazzi_2009,Geach_2009}.

Our results indicate that galaxy--galaxy interactions may be associated
with stronger or more recent truncated starbursts, and hence may be a
significant transition mechanism.  We therefore support the
conclusions of \citet{Moran_2007}, that a combination of 
galaxy--galaxy interactions, ram-pressure stripping, and other more
minor mechanisms are responsible for spiral to S0 transformation.

Galaxy--galaxy interactions have long been theoretically associated
with strong bar formation and nuclear starbursts (e.g.,
\citealt{Mihos_1996}).  However, due to the high relative velocities
of galaxies in a dense environment, tidal interactions can also have a
relatively gentle effect \citep{Moore_1996}. There is growing
observational evidence that even pair interactions may not cause
nuclear starbursts as readily as anticipated, enhancing star-formation
in spiral arms instead (e.g., \citealt{Casteels_2013}).  Furthermore,
bars are found to be prevalent in gas-poor, red spirals (e.g.,
\citealt{Masters_2011,Masters_2012}), and so may be more associated
with the suppression of star-formation, rather than its enhancement.

The final argument for a spiral to lenticular tranformation is the
properties of the final galaxies.  Lenticulars are consistent with
being formed from faded spirals in terms of their Tully-Fisher
relation \citep{Bedregal_2006}, globular cluster specific frequencies
\citep{Alfonso_2006}. However, they do tend to be more bulge
dominated \citep{Christlein_2004} and have hotter disks than spiral galaxies
\citep{Cortesi_2013}. This can be achieved by an enhancement of
central star-formation prior to transformation, and a marginal
increase in pressure support, perhaps through an accumulation of
galaxy-galaxy interactions. Both of these processes are suggested by
our results and many of the other studies discussed above.  The
clearing of dust in the central regions during the transition from
spiral to S0 may also enhance the bulge-to-disk ratio \citep{Driver_2007}.
Separately measuring the stellar population properties of bulges and
disks for large samples of spiral and S0 galaxies, in both
spectroscopic (e.g., \citealt{Johnston_2012}) and multi-band
photometric data (e.g.,
\citealt{Simard_2011,Lackner_2012,Bamford_2012,Haeussler_2013}),
will help to fill in many of the missing details.


\section{Conclusions}

The transformation from spiral galaxies into S0s, if it actually
occurs, must comprise a spectral transformation, resulting from the
suppression of star formation in the disk of the galaxy; a
morphological transformation, in terms of the removal of spiral features from the
disk and growth of the bulge; and a modest dynamical transformation,
with a small increase in the ratio of pressure versus rotational support.

We have studied the significance of disky `k+a' galaxies, indicative of a
spiral galaxy in which star formation was truncated
$\sim0.5$--$1.5$~Gyr ago, as the possible intermediate step in the
transformation of star-forming spirals into passive S0s in the
intermediate-redshift cluster environment.

These galaxies are typically identified by their strong Balmer
absorption line equivalent widths, an expected
signature of a dominant $\sim 1$ Gyr old stellar population.  We have
used spectral template fitting to show that galaxies selected via the
H$\delta_{\rm F}$ index do, indeed, contain significant fractions of
A-type stars and stellar populations with ages between $0.5$ and $1.5$
Gyr.  We study the spatial distribution of the
young population using these different indicators, finding generally
consistent results. While the disky `k+a' galaxies appear to be a
rather mixed population, their final episode of star-formation is
always distributed over a region of size similar to, or somewhat
smaller than, the older stars.

We have coarsely measured the velocity field of these galaxies, both in
terms of the full stellar population and, in a limited number of
cases, the separate young and old populations.  The results support
the picture that, in the majority of our sample, the last generation
of stars formed in a disk, in a very similar manner to previous
generations.

None of the disky `k+a' galaxies in this intermediate redshift cluster
appear to have experienced a violent event, such as a merger or
significant nuclear starburst, prior to the truncation of
their star-formation.  Instead, their regular disk star-formation has
simply ceased with only, in some cases, a small increase in central
concentration beforehand.

A relatively gentle mechanism must thus be responsible for the
cessation of star-formation.  Gas-related mechanisms, such as ram
pressure stripping, are therefore favoured.  However, there is also an
indication that many of our galaxies with more centrally concentrated
young populations have experienced recent galaxy-galaxy interactions.
This raises the possibility that, thanks to prior removal of the gas
halo, stochastic gravitational interactions may provide the necessary
impetus to halt star-formation, perhaps via a brief period of central
enhancement.

\section{Acknowledgments}

We thank Evelyn Johnston for her invaluable help modifying the
code and for very useful discussions. We also thank the 
anonymous referee for very useful comments that have helped improving 
the manuscript. BRDP acknowledges support from IAC and STFC.
BMJ acknowledges support from the ERC-StG grant EGGS-278202.
The Dark Cosmology Centre is funded by the Danish National Research
Foundation. Based on observations made with ESO telescopes at the La Silla 
Paranal Observatory under programme ID 073.A-0362A.

\bibliographystyle{mn2e}
\bibliography{my_ref}

\bsp

\label{lastpage}

\appendix
\section{Individual analysis of galaxies}

In this section we include the analysis of each of the
galaxies in the sample, with a qualitative description and 
the figures with individual analysis of each galaxy. 
In the figures we show the distribution of
light in the individual IFUs, the integrated spectra of the
galaxies, the distribution of the three different indicators
H$\delta_{\rm F}$, A/(AFGKM) and $f_{\rm young}$ throughout
the galaxies, as well as the maps of velocity and $\sigma$ for the
whole galaxy and for the old and young populations, as obtained
with the simultaneous fitting procedure.

\subsubsection{CN4}

This elliptical galaxy has low values of H$\delta$$_{\rm F}$ and
$f_{\rm young}$, therefore being possibly misclassified as `k+a'
galaxy by CS87.  However, there seems to be a relatively important population
of B and A stars in the stellar template histograms.  No pattern of
rotation was found in the kinematic analysis. It is isolated.

\subsubsection{CN22}

This galaxy, which is classified as peculiar, seems to be 
an ongoing merger from inspection of the HST/WFPC2 images. Its distributions of
A/(AFGKM) and $f_{\rm young}$ are consistent with the young population
been concentrated in the centre of the galaxy, although the H$\delta$$_{\rm F}$ 
maps show a more extended distribution. The value of H$\delta$$_{\rm F} = 6.4
\pm 0.4$~\AA{} implies the occurrence of a starburst to produce the
k+a feature and not a simple truncation of the star formation in the
galaxy. The starburst may have taken place in the centre of the galaxy
although the distribution of H$\delta$$_{\rm F}$ implies a more
extended young population. Rotation was found in this galaxy together
with higher values of $\sigma$ in the centre. The two populations
found in the kinematic decomposition are rotating in the same
direction and both show higher values of $\sigma$ in the centre.

The merger appears to be responsible for producing a
centrally-concentrated young stellar population
before halting star-formation, resulting in the `k+a' spectrum
observed.

\subsubsection{CN24}

This galaxy has low global values for the three young population
indicators. In particular H$\delta_{\rm F} = 1.0 \pm 0.4$~\AA{} and
therefore it appears to have been misclassified as `k+a' by CS87. The
velocity maps do not show a clear pattern of rotation. The galaxy is
isolated. CN24 is consistent with being a passive spiral galaxy.

\subsubsection{CN74}

With low values of all the young population indicators, this galaxy
also appears to have an unreliable H$\delta$ measurement by CS87.  We
measure a global value of H$\delta_{\rm F} = 1.0 \pm 0.3$~\AA.  CS87
detect emission in [OII], although there were no emission features
found in our spectra of the galaxy. It presents clear rotation. 
This galaxy does not show signs of recent interaction and is isolated.

\subsubsection{CN89}

This elliptical galaxy has very low values of A/(AFGKM) and $f_{\rm
  young}$ although its global H$\delta_{\rm F} = 2.1 \pm 0.9$~\AA,
showing a uniform distribution of the young population. No kinematic analysis could be performed for
this galaxy.

\subsubsection{CN119}

The global values of the young population indicators in this galaxy
are low, with H$\delta_{\rm F} = 1.0 \pm 0.5$~\AA. However, the value
of $f_{\rm young} = 0.4$ does suggest the presence of a young
population in the galaxy that is not dominant enough to present clear
spectral features.

The weights of the different SSP templates suggest that there are two
dominant populations, one with ages between $0.5$ and $1.5$~Gyr and
the other one older than $7$~Gyr, which are also found in the
kinematic decomposition. Both populations have similar velocity
distributions although the old population show generally higher values
of sigma. Although this galaxy would be consistent with a rotating
system, the analysis of the individual IFU elements did not provide
enough information for its confirmation. The galaxy is not
interacting.

\subsubsection{CN143}

Very high values of all the young population indicators, imply the occurrence of 
a recent ($< 1.5$~Gyr) starburst in the galaxy. The distribution maps show the young population
extended throughout the whole galaxy. Rotation is detected in this galaxy although
no kinematic decomposition could be performed. The galaxy is isolated.

These findings are consistent with this system being a spiral galaxy where the gas has 
been depleted and used up in a starburst. Because there are no signs of interaction, this
depletion may have been due to the interaction with the intracluster medium.

\subsubsection{CN146}

This galaxy has detected emission in [OII] by CS87, and emission lines
can be seen in its Balmer absorption features in our spectrum. The
values of the indicators are consistent with the presence of a young
population in the galaxy. This population appears spatially extended
in the galaxy although the $f_{\rm young}$ maps show higher
concentration in the centre. No clear pattern of rotation or trend of
sigma is found in the maps of the kinematics. Alhough there are few
individual IFU elements with good S/N, the kinematic decomposition
shows higher values of sigma in the outskirts for the young population
whereas the old population has higher velocity dispersions in the
centre. This galaxy is not interacting.

The characteristics of this galaxy imply star formation that is
gradually declining but has not been entirely truncated yet,
consistent with the depletion of gas due to interaction with the ICM.

\subsubsection{CN155}

This galaxy has [OII] emission detected by CS87, and emission lines
can be seen in its Balmer absorption features in our spectrum.  The
distribution of the young population indicators is consistent with the
young population being more dominant in the central regions. The
strong value of H$\delta_{\rm F}$ implies the occurrence of a
starburst. No clear pattern of rotation was found and no kinematic
decomposition could be performed in this galaxy. In the HST/WFPC2
image this galaxy is interacting with a smaller object. The
post-starburst feature may be associated with an interaction with this
companion. The star formation in the galaxy has not been truncated
yet, therefore this galaxy could be similar to the progenitors of the
k+a galaxies in our sample.

\subsubsection{CN187}

The global values of the young population indicators in this galaxy
are low, with H$\delta_{\rm F} = 1.0 \pm 0.5$~\AA.  The kinematics of
this galaxy could not be analysed due to the low number of IFU
elements available and their distribution. This galaxy is
isolated. This galaxy appears to have been misclassified as a `k+a'
galaxy by CS87.

\subsubsection{CN191}

This galaxy presents high values of all the young population
indicators, with H$\delta_{\rm F} = 5.1 \pm 0.3$~\AA, showing flat
distributions. Although the H$\delta_{\rm F}$ value is not
exceptionally high, it could be consistent with the occurrence of a
starburst in the galaxy between $0.5$ and $1.5$~Gyr ago.  The galaxy
shows clear rotation and its kinematical decomposition shows two
populations (young and old) rotating in the same direction, with the
young population having higher values of sigma in the centre. This
galaxy is not found to be interacting.

The presence of rotation implyes that the process responsible for the
truncation of the star formation did not affect the kinematic state of
the galaxy. The distribution of the young population implies a that
the final episode of star formation occured throughout the galaxy.

\subsubsection{CN228}

The young population in this galaxy is concentrated in the central
regions with high global values of the indicators.  It displays
rotation and the two distinct stellar populations are rotating in the
same direction. It has a very close satellite.

The presence of a close satellite and the distribution of the young
population suggest the interaction with the other object as the
responsible mechanism for the truncation of the star formation, which
has not affected the kinematics of the galaxy.

\subsubsection{CN229}

This galaxy, which was observed instead of CN254 by P05, is a disk
system which has very low values of H$\delta_{\rm F} = 0.1 \pm
1.1$~\AA, but whose histograms show the presence of a very young `O'
stars, and a population with age $< 0.5$~Gyr. Although no [OII] has
been measured in this galaxy, this dominant young population would be
consistent with ongoing star formation. This galaxy therefore does not
fulfill the `k+a' criteria, nor does it show evidence of rapidly
declining star-fomation.

\subsubsection{CN232}

Very low global values of all the indicators, with H$\delta_{\rm F} =
1.1 \pm 0.6$~\AA{} imply there is no dominant young population. This
galaxy presents rotation but it could not be kinematically
decomposed. In the images it is found with a close satellite. This
galaxy seems to have been misclassified by CS87 and is not an actual
k+a.

\subsubsection{CN243}

This galaxy has two broken fibres in the centre, and therefore no
spatial analysis could be performed.  Its has a global value
of H$\delta_{\rm F} = 3.1 \pm 0.3$~\AA{} and it CS87 detected emission
in OII.  Due to the broken fibres we could not obtain maps of the
kinematics. This galaxy is in a close encounter with other object,
with which it seems to be exchanging material.

\subsubsection{CN247}

An elliptical galaxy with very low values of all the indicators. One
fibre has H$\delta_{\rm F} \sim 3.0$~\AA, although the global value is
much lower at $1.0 \pm 0.4$~\AA. No kinematic analysis could be
performed for this galaxy. Its characteristics are consistent with an
old, passive galaxy.
 
\subsubsection{CN254}

Although this galaxy was meant to be observed by P05 as well as us,
their observation actually corresponded to galaxy CN229. H$\delta_{\rm
  F}$ is the only young population indicator with high values and it
shows similar values throughout the galaxy, with a global value of $2.1 \pm 0.5$~\AA.  This
galaxy shows rotation and and two distinct populations that are
rotating in the same direction. The old population shows higher values
of sigma than the young in the whole galaxy. This galaxy is surrounded
by smaller objects but does not show signs of interaction.

The values of H$\delta_{\rm F}$, the presence of rotation in
both populations and the fact that the galaxy is not interacting
suggest that the truncation of the star formation was produced by the
gradual removal of the gas in the disk of the galaxy due to an
interaction with the ICM.

\subsubsection{CN667}

Disk galaxy with low values of A/(AFGKM) and $f_{\rm young}$ and very
low H$\delta_{\rm F} = 0.6 \pm 0.4$~\AA. No kinematic analysis could
be performed in this galaxy. This galaxy is possibly interacting with
two close satellites.

\subsubsection{CN849}

A disk galaxy with relatively low values of the young population
indicators, with H$\delta_{\rm F} = 2.3 \pm 0.4$~\AA. The distribution of the indicators
suggest a concentration of the young population in the outskirts, but
H$\delta_{\rm F}$ is also prominent in the central pixel. The
histograms of the SSP models show two populations, one young and one
old, which are also found in the kinematic decomposition to be
rotating in the same direction, although the young component appears
to rotate faster. This galaxy is clearly exchanging material with
another object, and therefore this interaction could be responsible
for the cessation of star formation.

\subsubsection{CN858}

This elliptical galaxy has a zero value of A/(AFGKM) and $f_{\rm
  young}$ and negative H$\delta_{\rm F}$, implying that it is a
passive galaxy. It is not interacting.

\clearpage
\newpage
\begin{figure*}
\begin{center}
\hspace{0cm}
\vspace{0.2cm}
\LARGE{CN4}

\includegraphics[height=0.88\textheight]{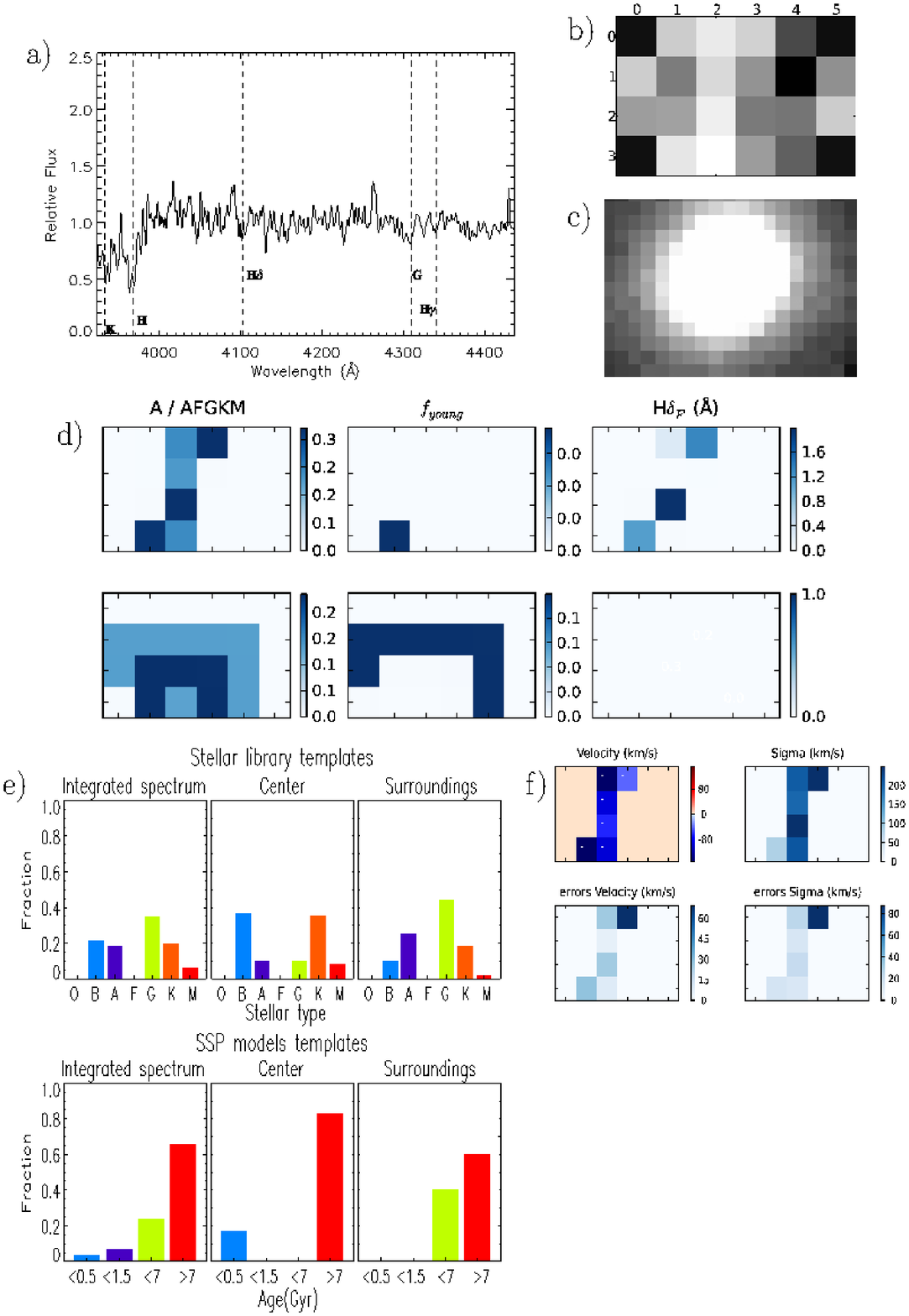}

\caption{Individual analysis of each galaxy. (a) Integrated spectra, (b) distribution of light in the IFU, 
(c) HST/WFPC2 image of the galaxy, (d) maps of the individual fibre values of A/(AFGKM), $f_{\rm young}$ and
 H$\delta_{\rm F}$ index (top) and the corresponding values for the integrated regions
`centre', `surroundings' and `outskirts' when available (bottom). Errors of the H$\delta_{\rm F}$
 index are printed over the regions. Each spatial pixel has a size of 0.52 x 0.52 which
 corresponds to $\sim$ 2.3 x 2.3 kpc$^{2}$ at the redshift of AC114. 
(e) Histograms of stellar type and stellar population age obtained with pPXF for the integrated 
spectra, `centre', `surroundings' and `outskirts' when available. (f) Maps of velocity and $\sigma$ for the galaxy. 
(g) Maps of velocity and sigma for the old and young populations (not available for CN4). Blank spaces are left when the respective analysis
could not be performed in a galaxy.} 
\end{center}
\end{figure*}

\begin{figure*}
\begin{center}
\LARGE{CN22}
\hspace{0cm}
\vspace{0.2cm}
\includegraphics[height=0.98\textheight]{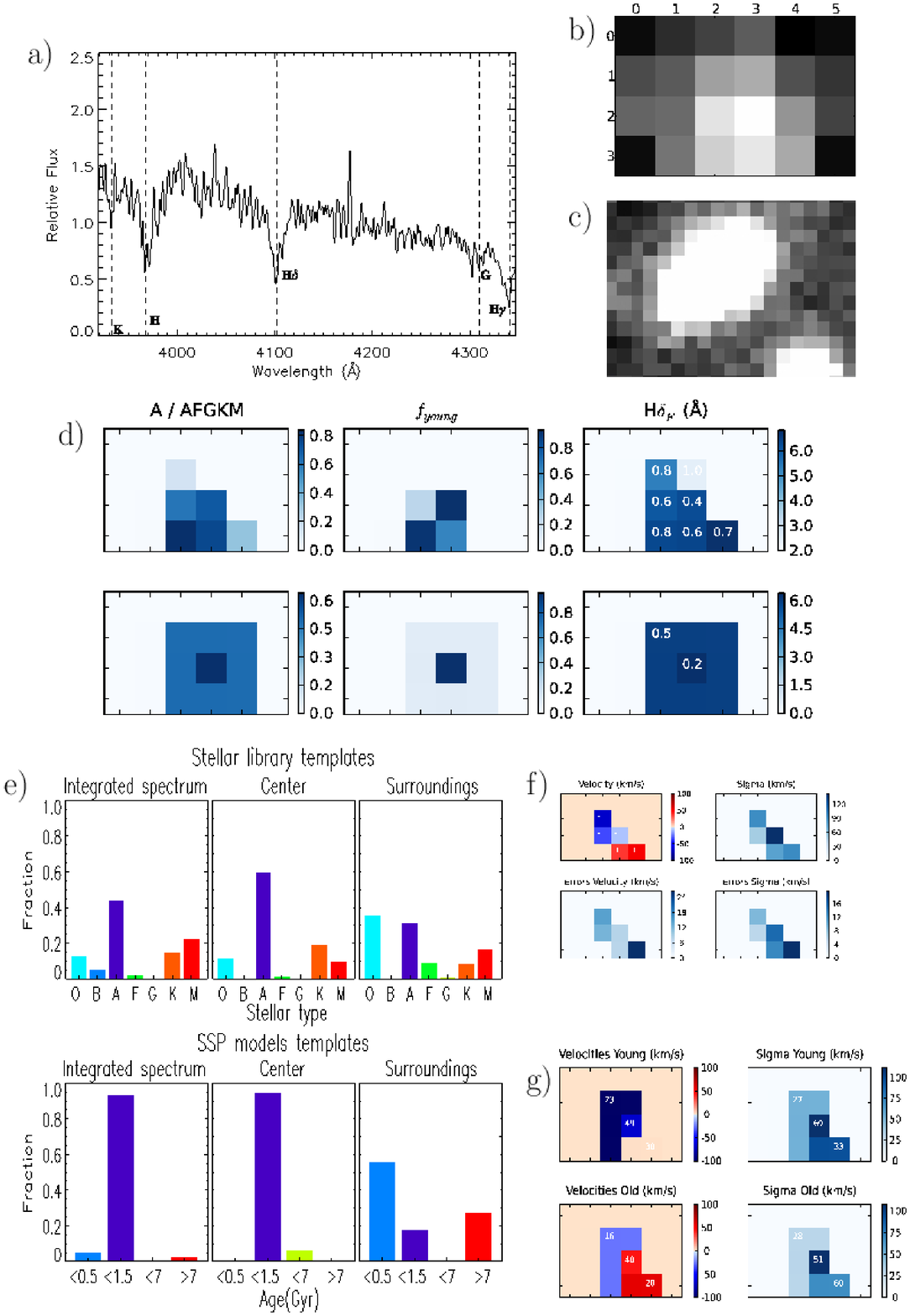}
\end{center}
\end{figure*}

\begin{figure*}
\begin{center}
\LARGE{CN24}
\hspace{0cm}
\vspace{0.2cm}
\includegraphics[height=0.98\textheight]{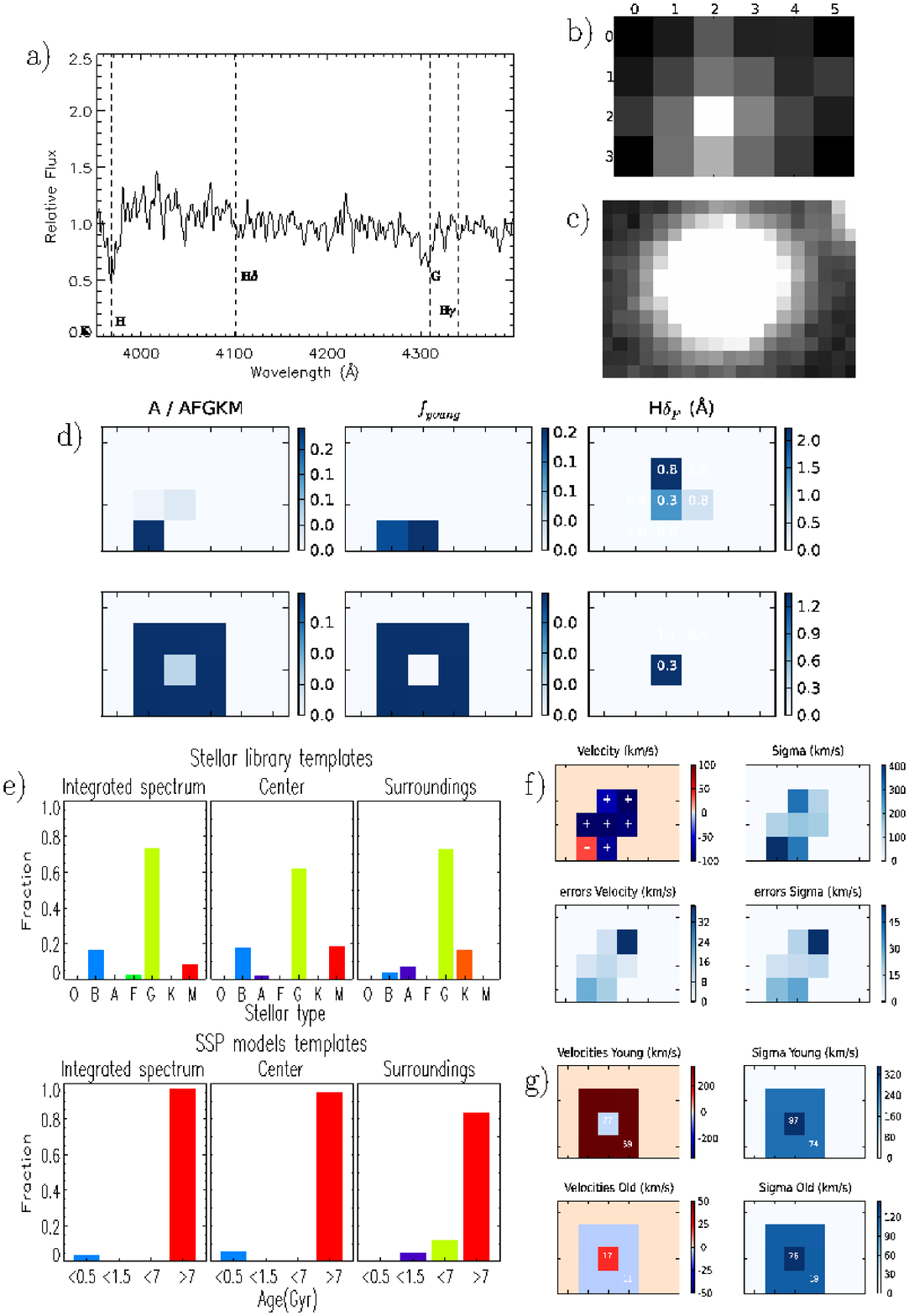}
\end{center}
\end{figure*}

\begin{figure*}
\begin{center}
\LARGE{CN74}
\hspace{0cm}
\vspace{0.2cm}
\includegraphics[height=0.98\textheight]{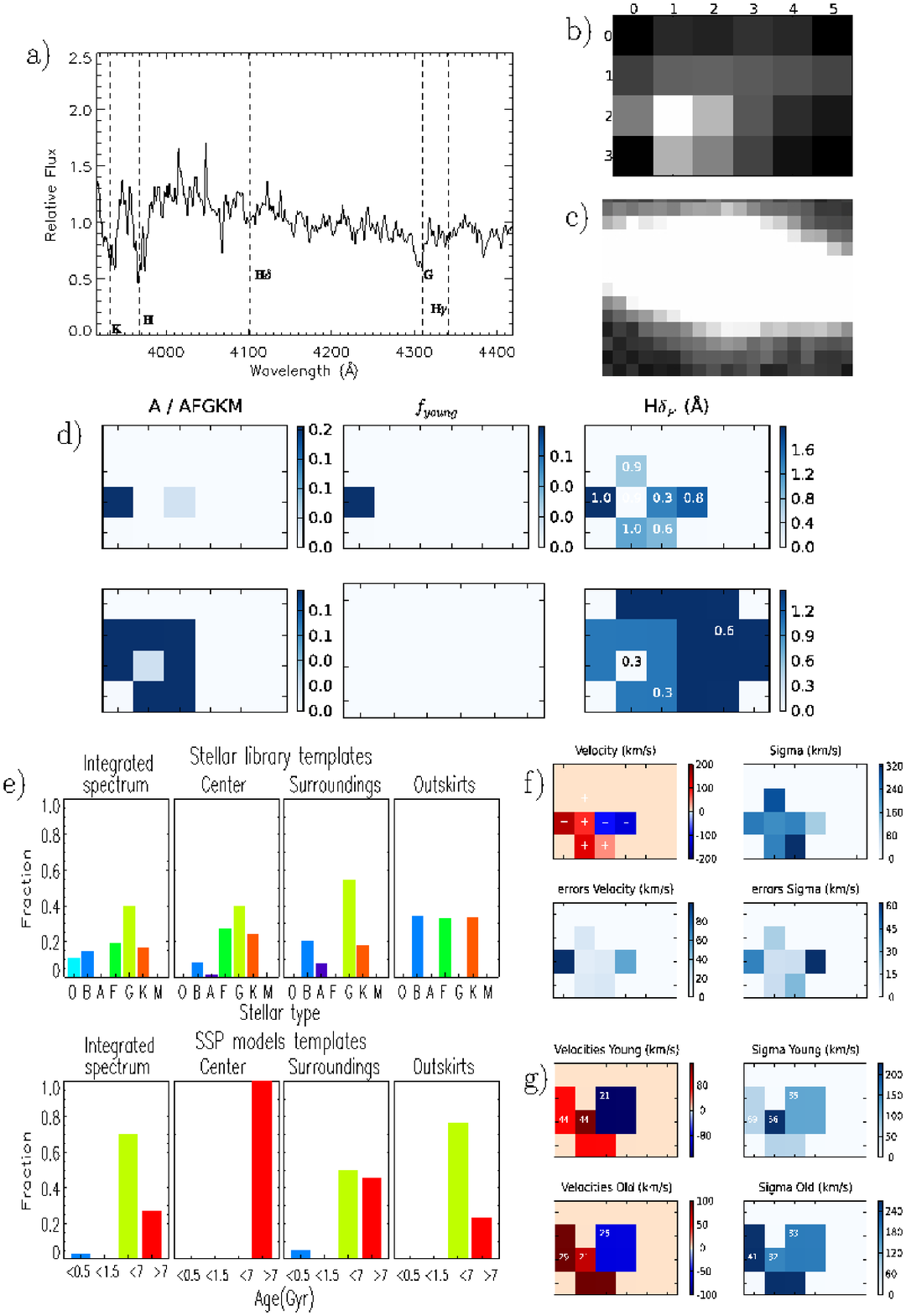}
\end{center}
\end{figure*}

\begin{figure*}
\begin{center}
\LARGE{CN89}
\hspace{0cm}
\vspace{0.2cm}
\includegraphics[height=0.98\textheight]{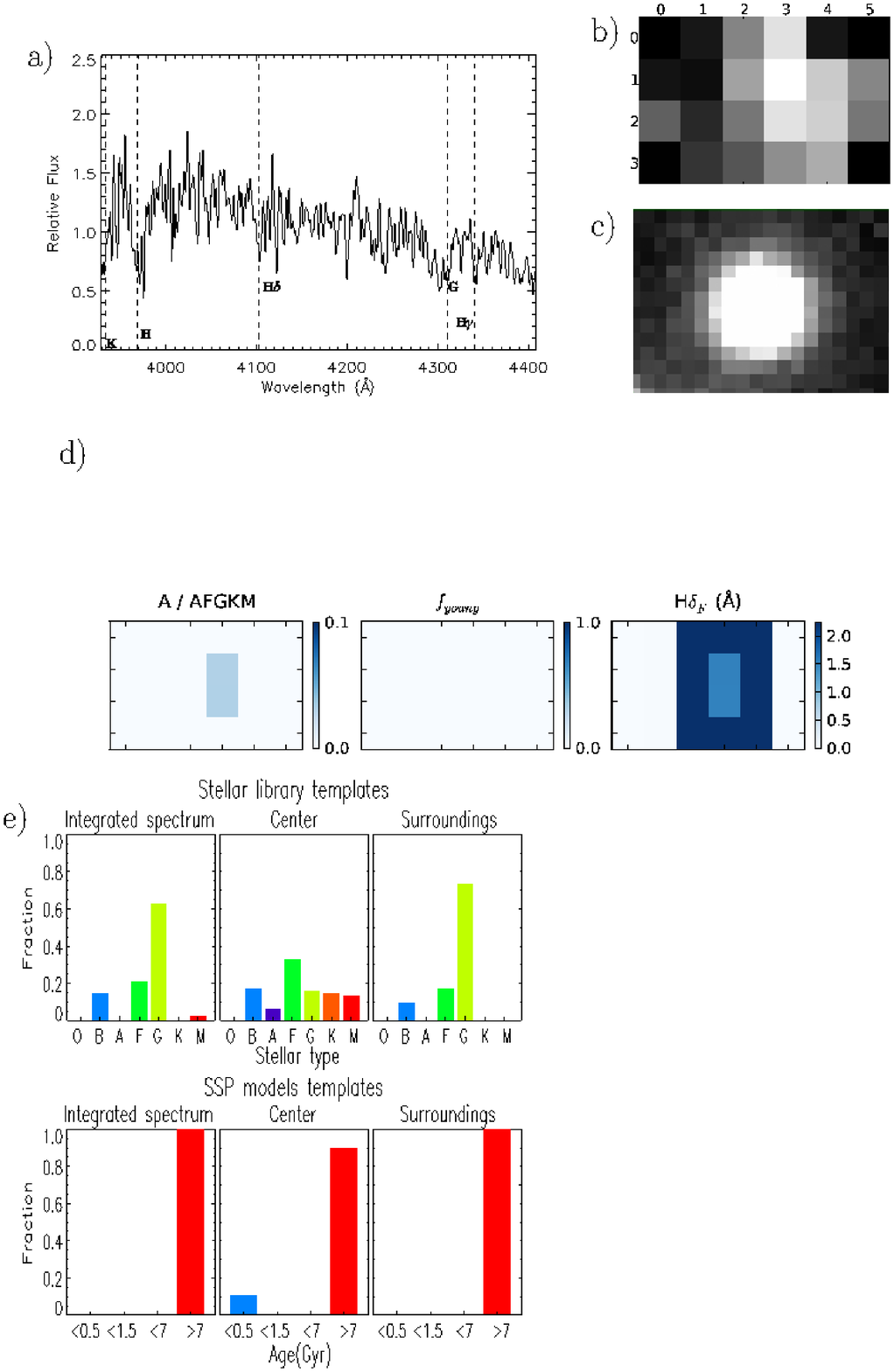}
\end{center}
\end{figure*}

\begin{figure*}
\begin{center}
\LARGE{CN119}
\hspace{0cm}
\vspace{0.2cm}
\includegraphics[height=0.98\textheight]{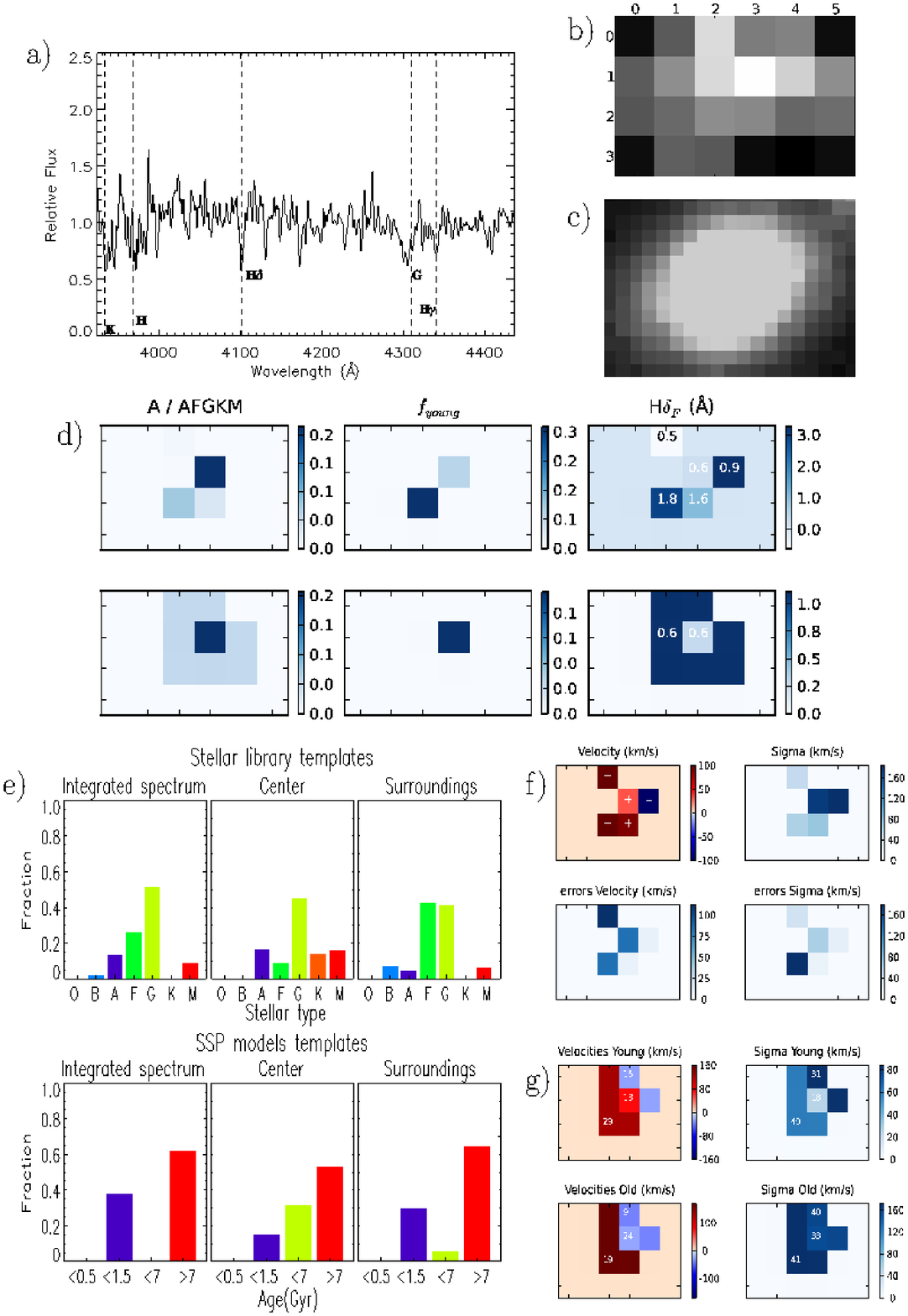}
\end{center}
\end{figure*}

\begin{figure*}
\begin{center}
\LARGE{CN143}
\hspace{0cm}
\vspace{0.2cm}
\includegraphics[height=0.98\textheight]{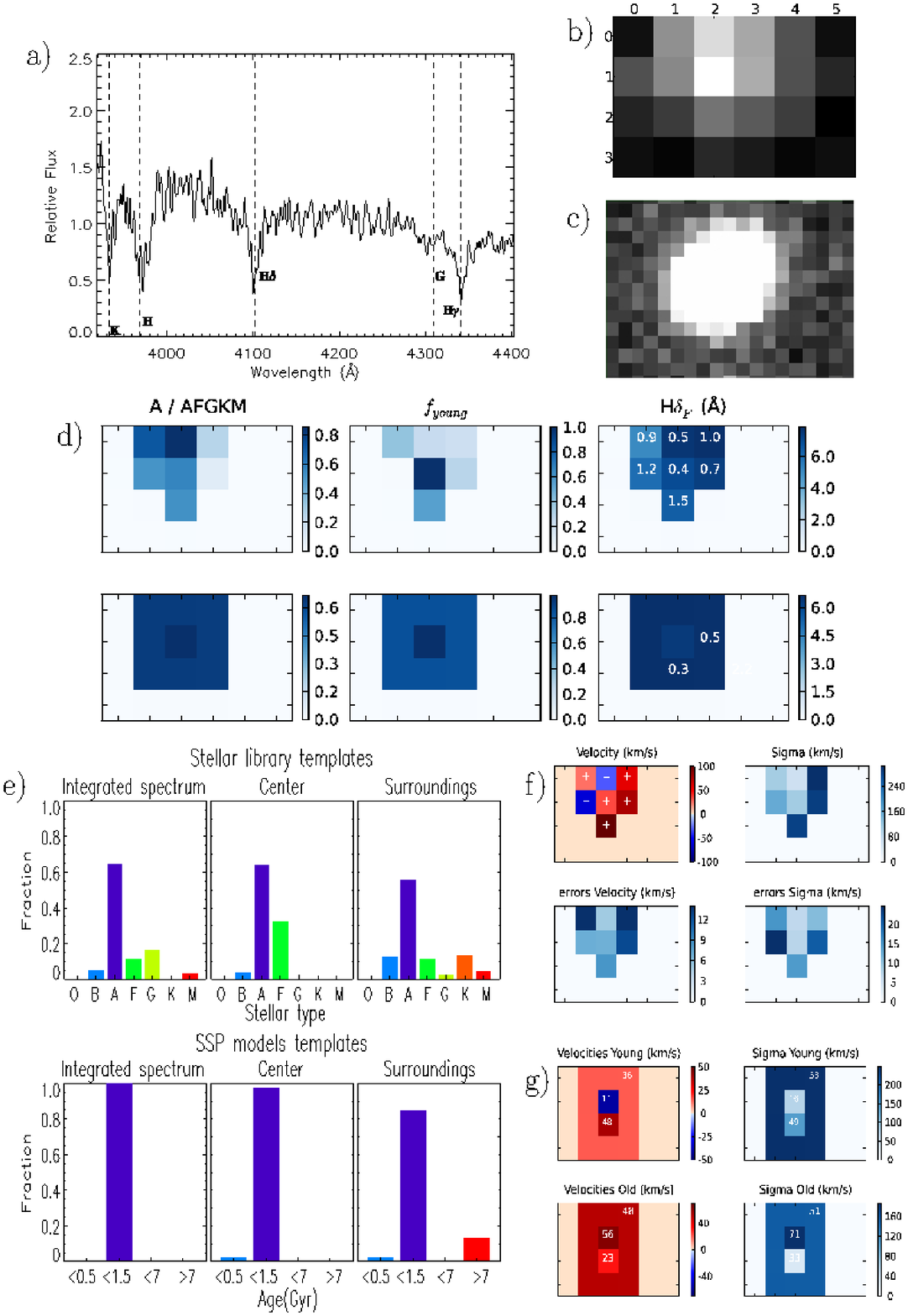}
\end{center}
\end{figure*}

\begin{figure*}
\begin{center}
\LARGE{CN146}
\hspace{0cm}
\vspace{0.2cm}
\includegraphics[height=0.98\textheight]{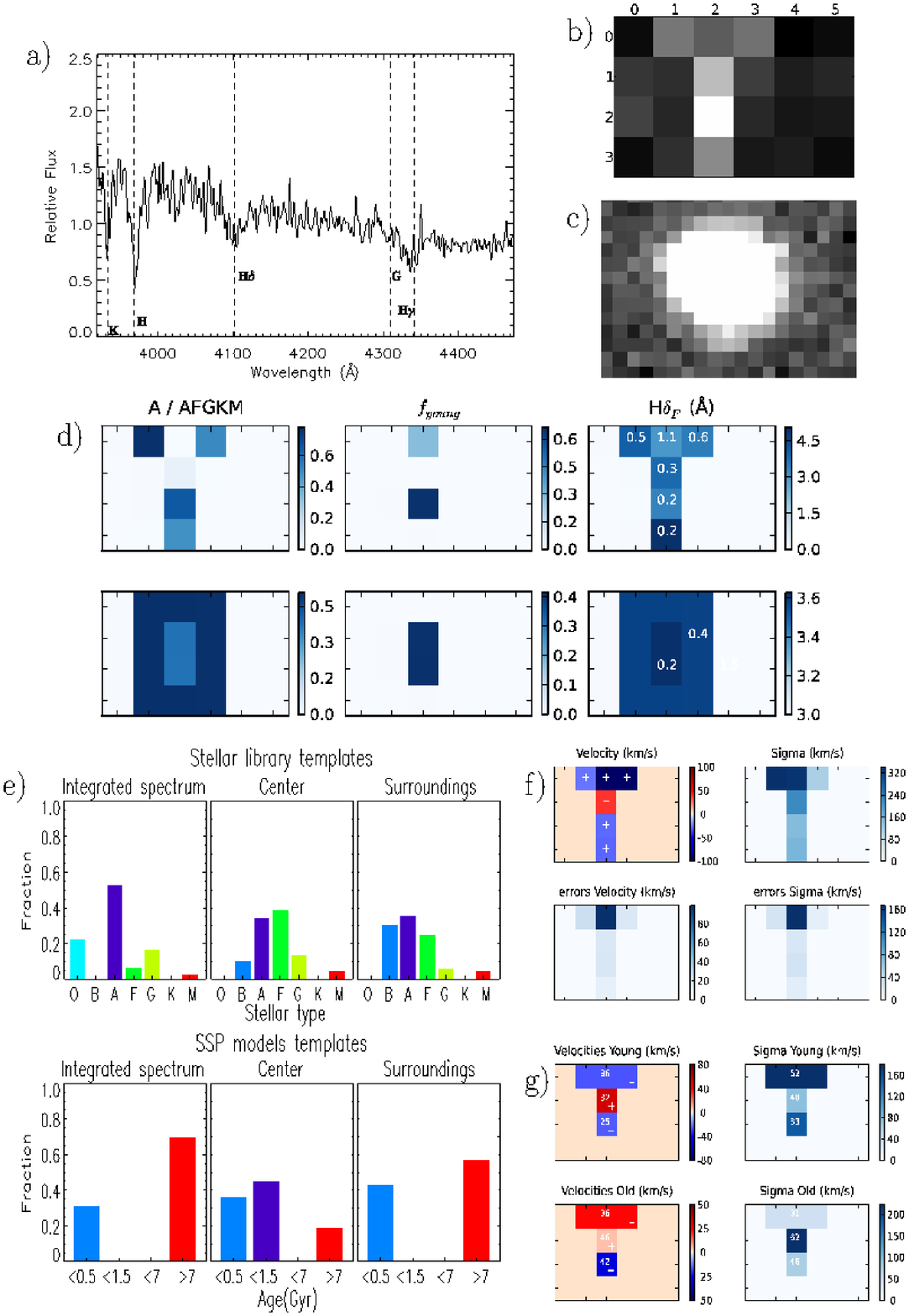}
\end{center}
\end{figure*}

\begin{figure*}
\begin{center}
\LARGE{CN155}
\hspace{0cm}
\vspace{0.2cm}
\includegraphics[height=0.98\textheight]{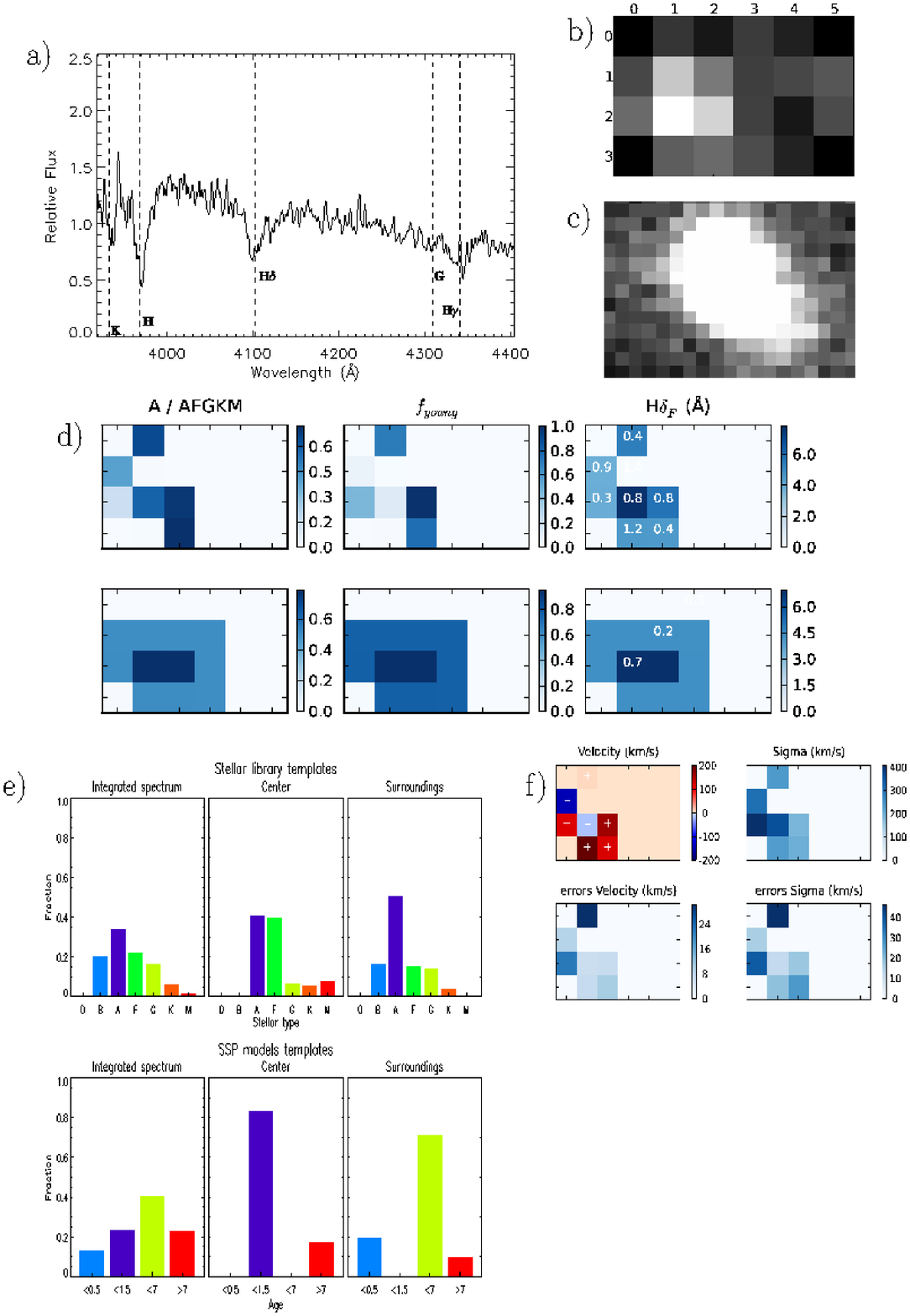}
\end{center}
\end{figure*}

\begin{figure*}
\begin{center}
\LARGE{CN187}
\hspace{0cm}
\vspace{0.2cm}
\includegraphics[height=0.98\textheight]{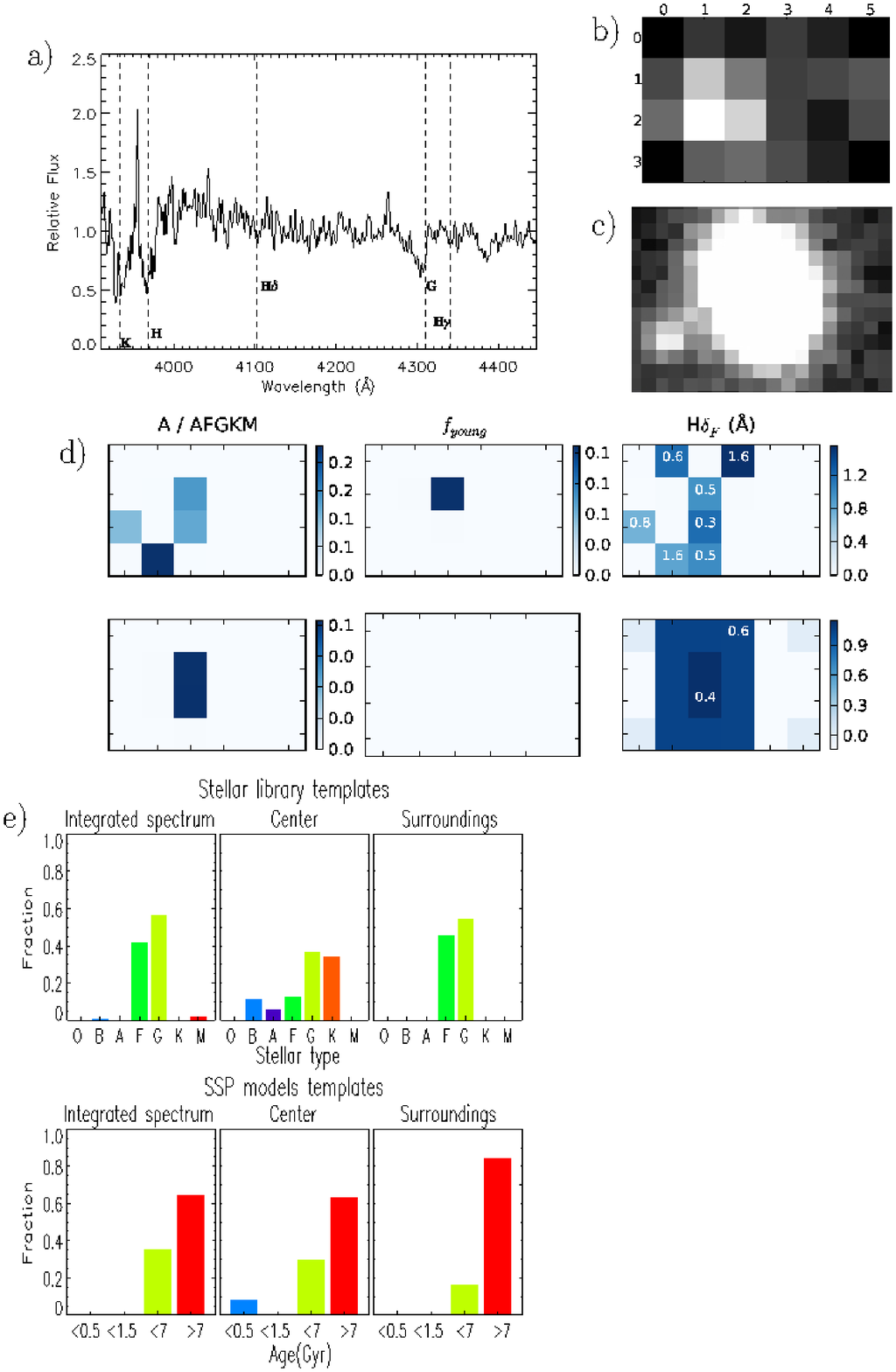}
\end{center}
\end{figure*}

\begin{figure*}
\begin{center}
\LARGE{CN191}
\hspace{0cm}
\vspace{0.2cm}
\includegraphics[height=0.98\textheight]{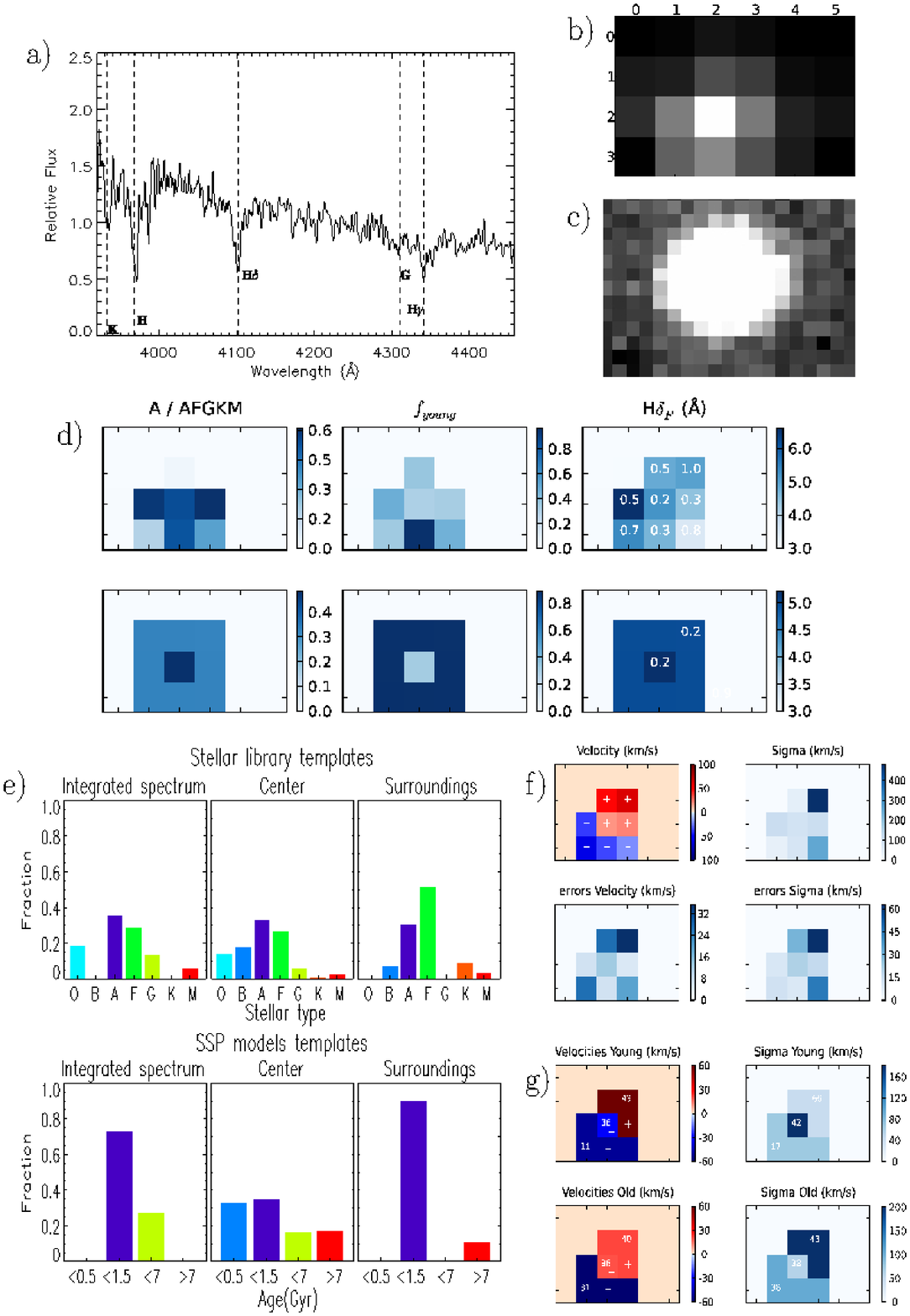}
\end{center}
\end{figure*}

\begin{figure*}
\begin{center}
\LARGE{CN228}
\hspace{0cm}
\vspace{0.2cm}
\includegraphics[height=0.98\textheight]{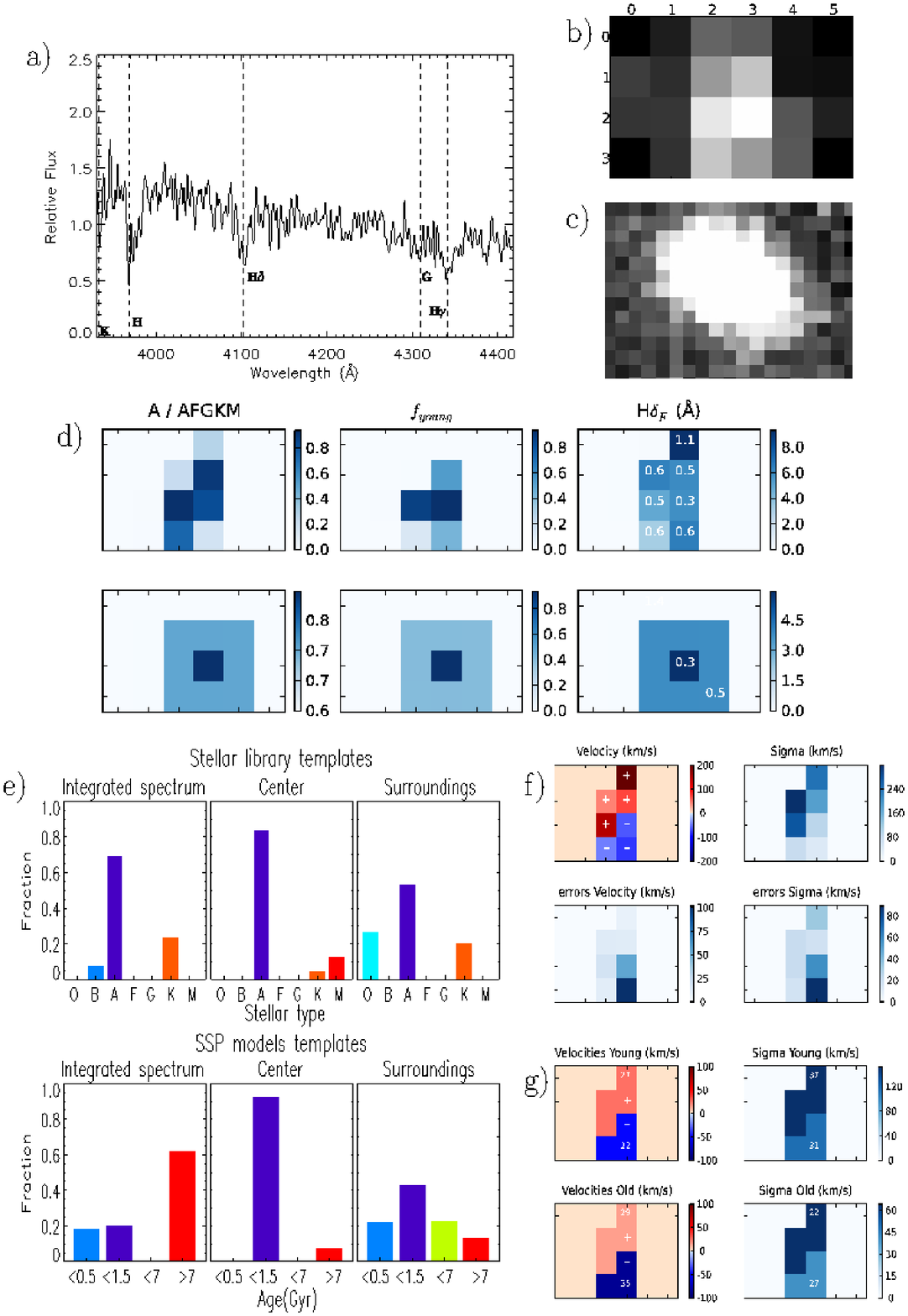}
\end{center}
\end{figure*}

\begin{figure*}
\begin{center}
\LARGE{CN229}
\hspace{0cm}
\vspace{0.2cm}
\includegraphics[height=0.98\textheight]{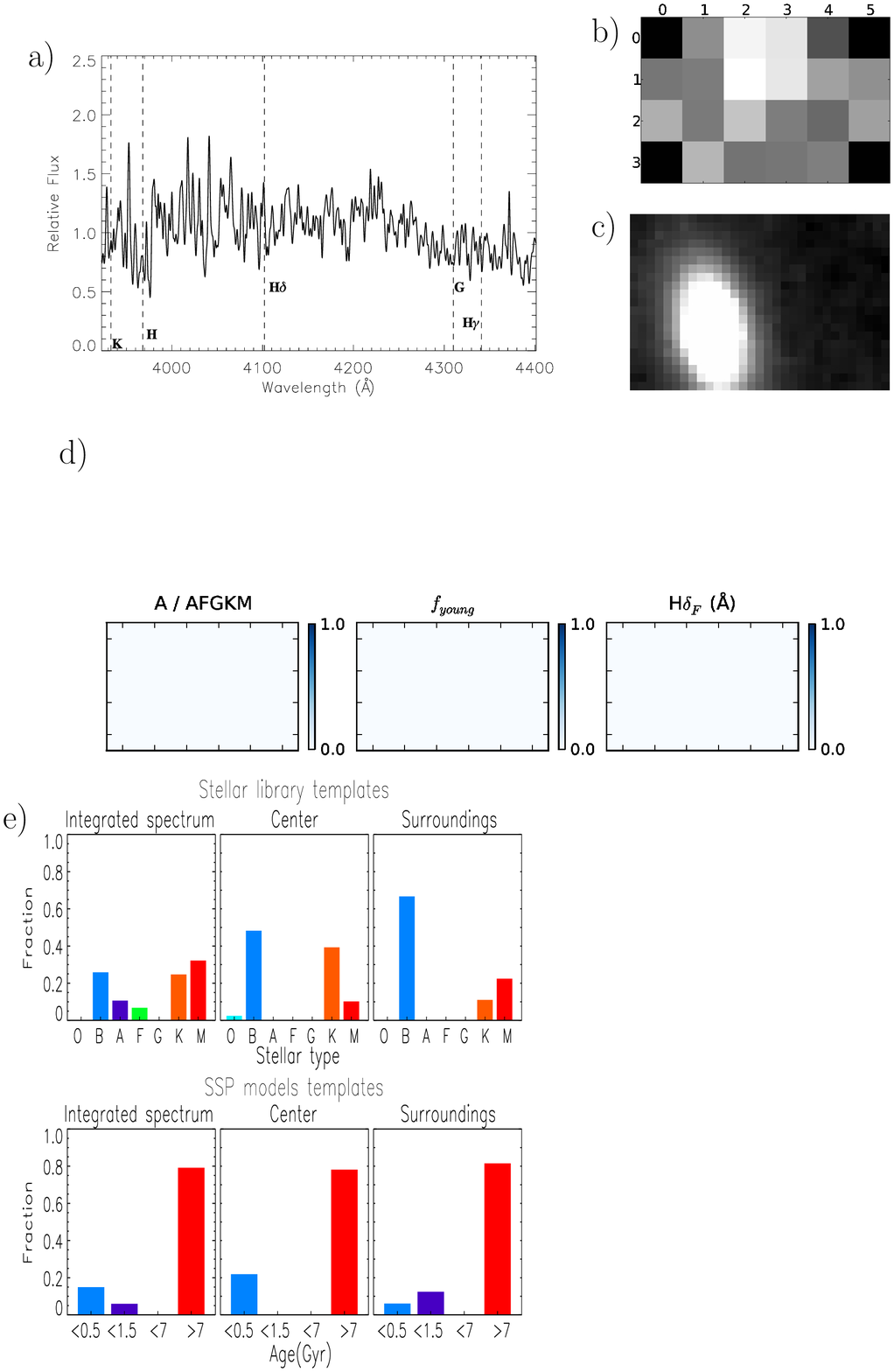}
\end{center}
\end{figure*}

\begin{figure*}
\begin{center}
\LARGE{CN232}
\hspace{0cm}
\vspace{0.2cm}
\includegraphics[height=0.98\textheight]{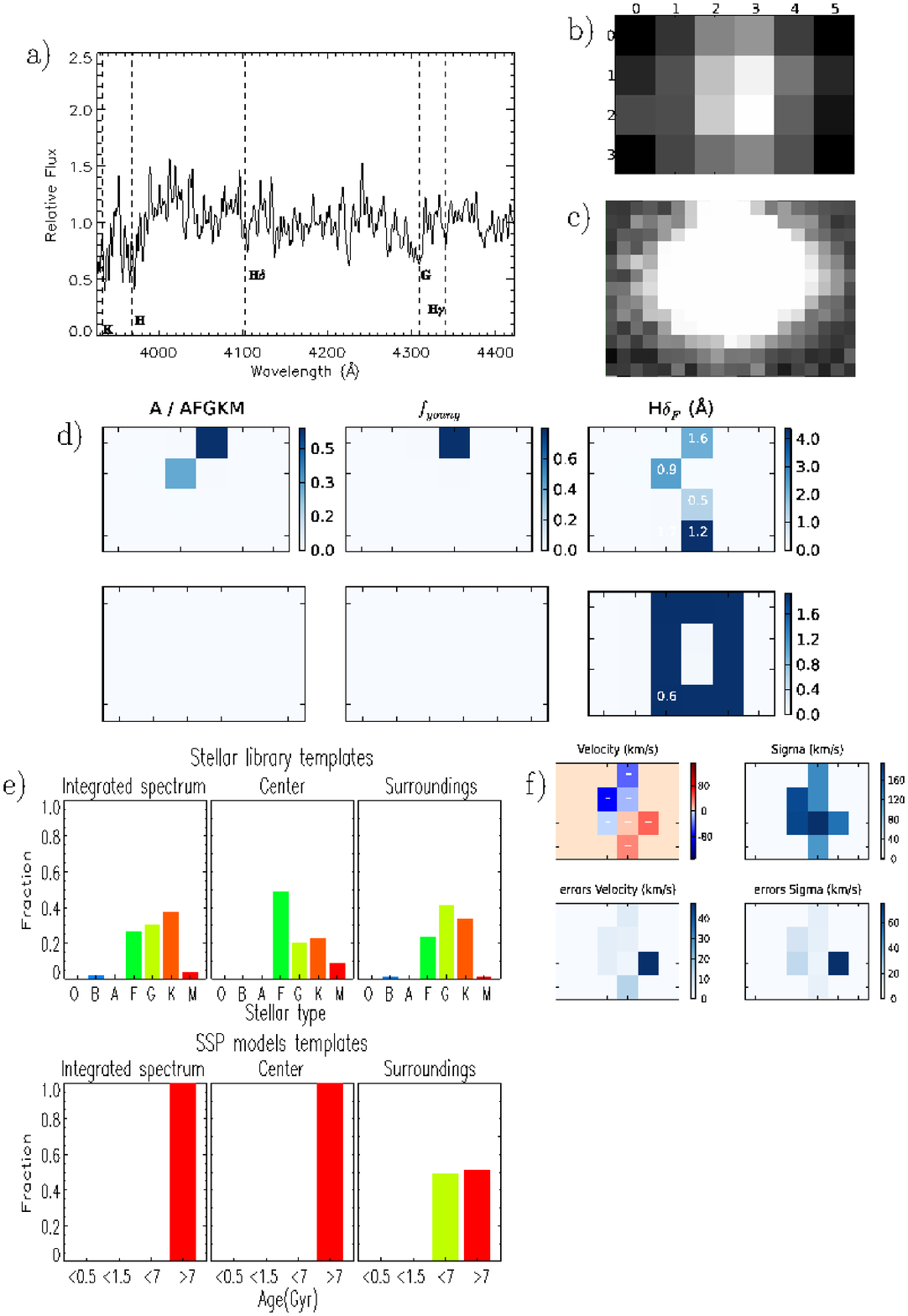}
\end{center}
\end{figure*}

\begin{figure*}
\begin{center}
\LARGE{CN243}
\hspace{0cm}
\vspace{0.2cm}
\includegraphics[height=0.98\textheight]{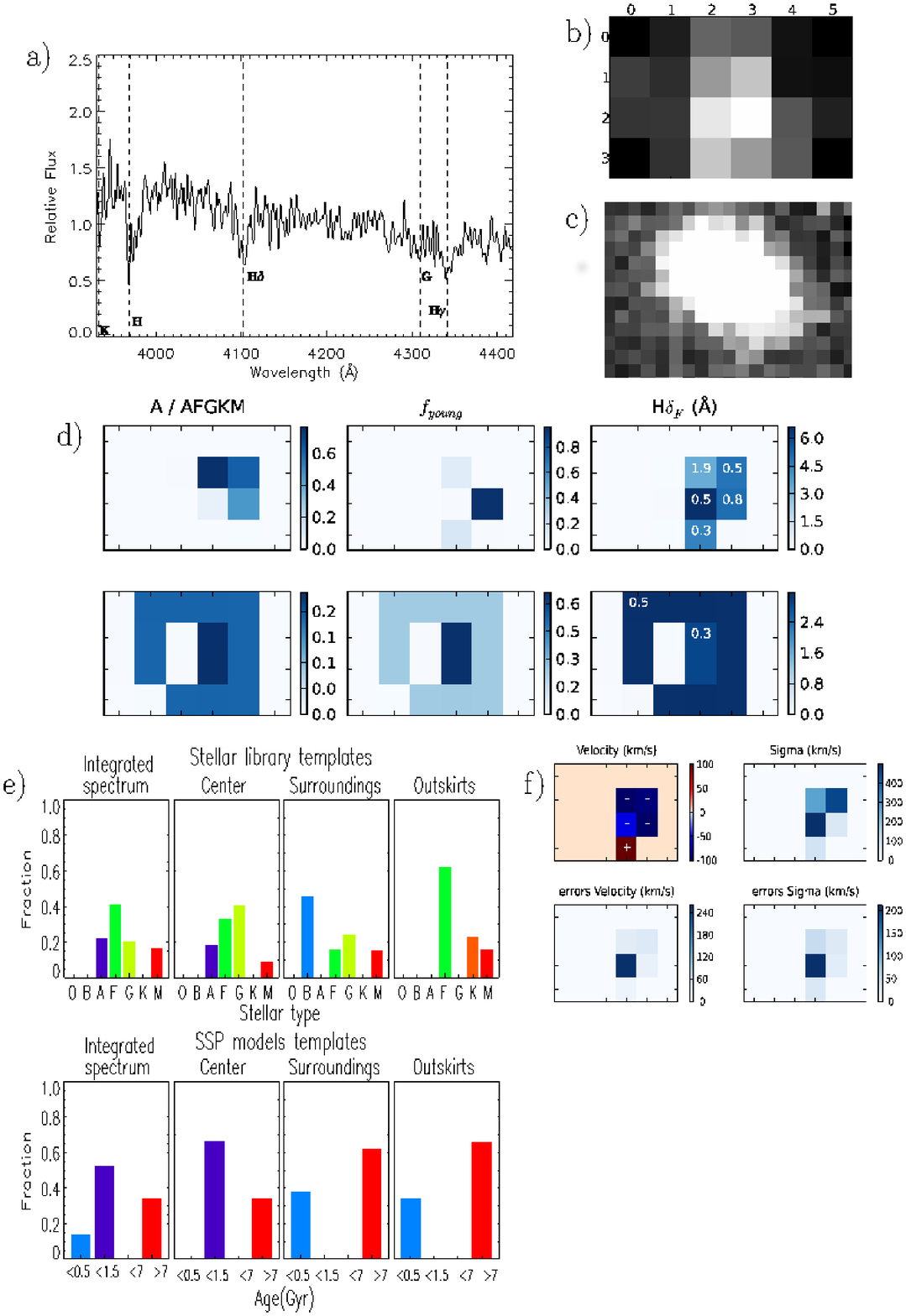}
\end{center}
\end{figure*}

\begin{figure*}
\begin{center}
\LARGE{CN247}
\hspace{0cm}
\vspace{0.2cm}
\includegraphics[height=0.98\textheight]{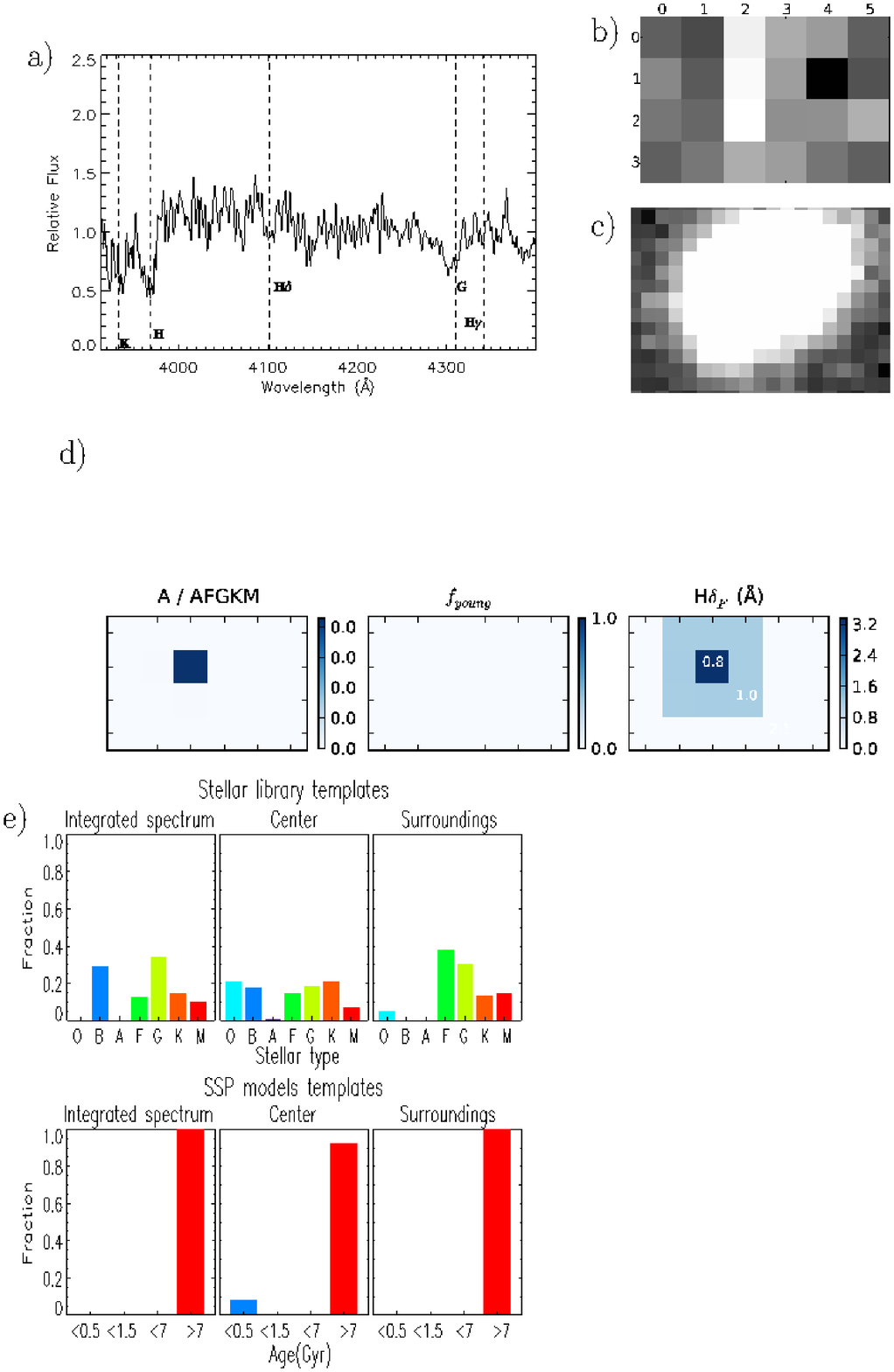}
\end{center}
\end{figure*}

\begin{figure*}
\begin{center}
\LARGE{CN254}
\hspace{0cm}
\vspace{0.2cm}
\includegraphics[height=0.98\textheight]{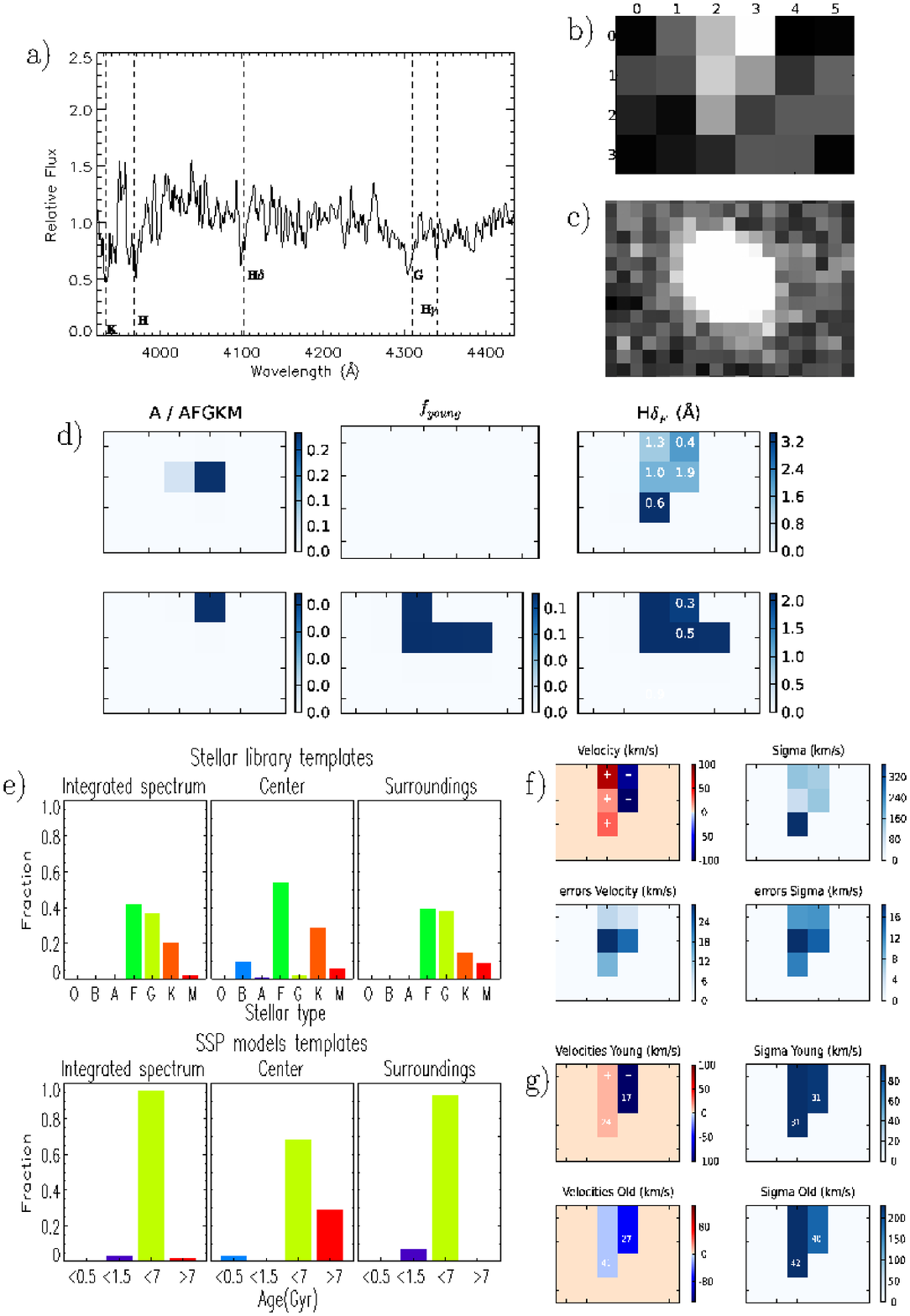}
\end{center}
\end{figure*}
\clearpage

\begin{figure*}
\begin{center}
\LARGE{CN667}
\hspace{0cm}
\vspace{0.2cm}
\includegraphics[height=0.98\textheight]{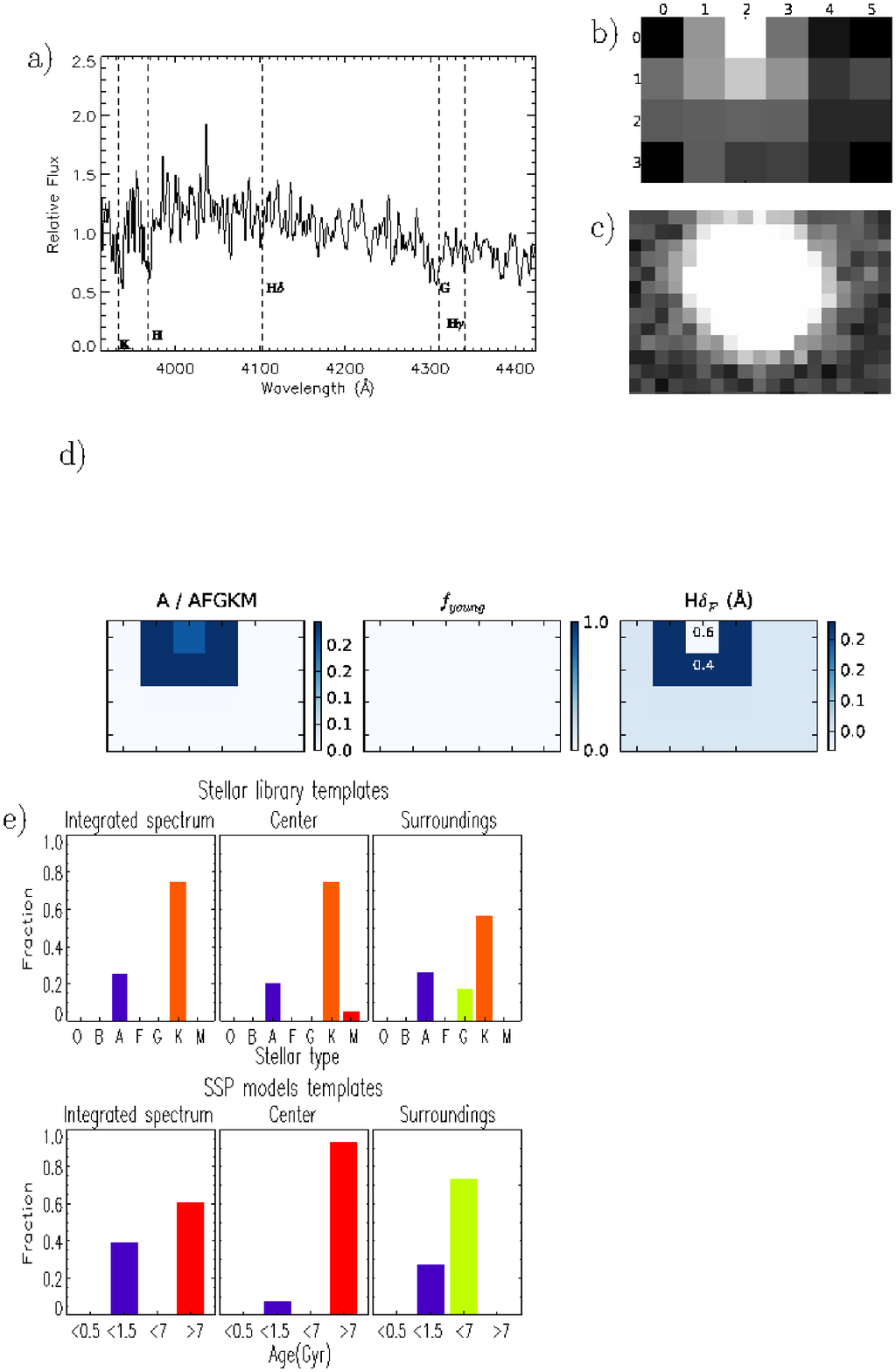}
\end{center}
\end{figure*}

\begin{figure*}
\begin{center}
\LARGE{CN849}
\hspace{0cm}
\vspace{0.2cm}
\includegraphics[height=0.98\textheight]{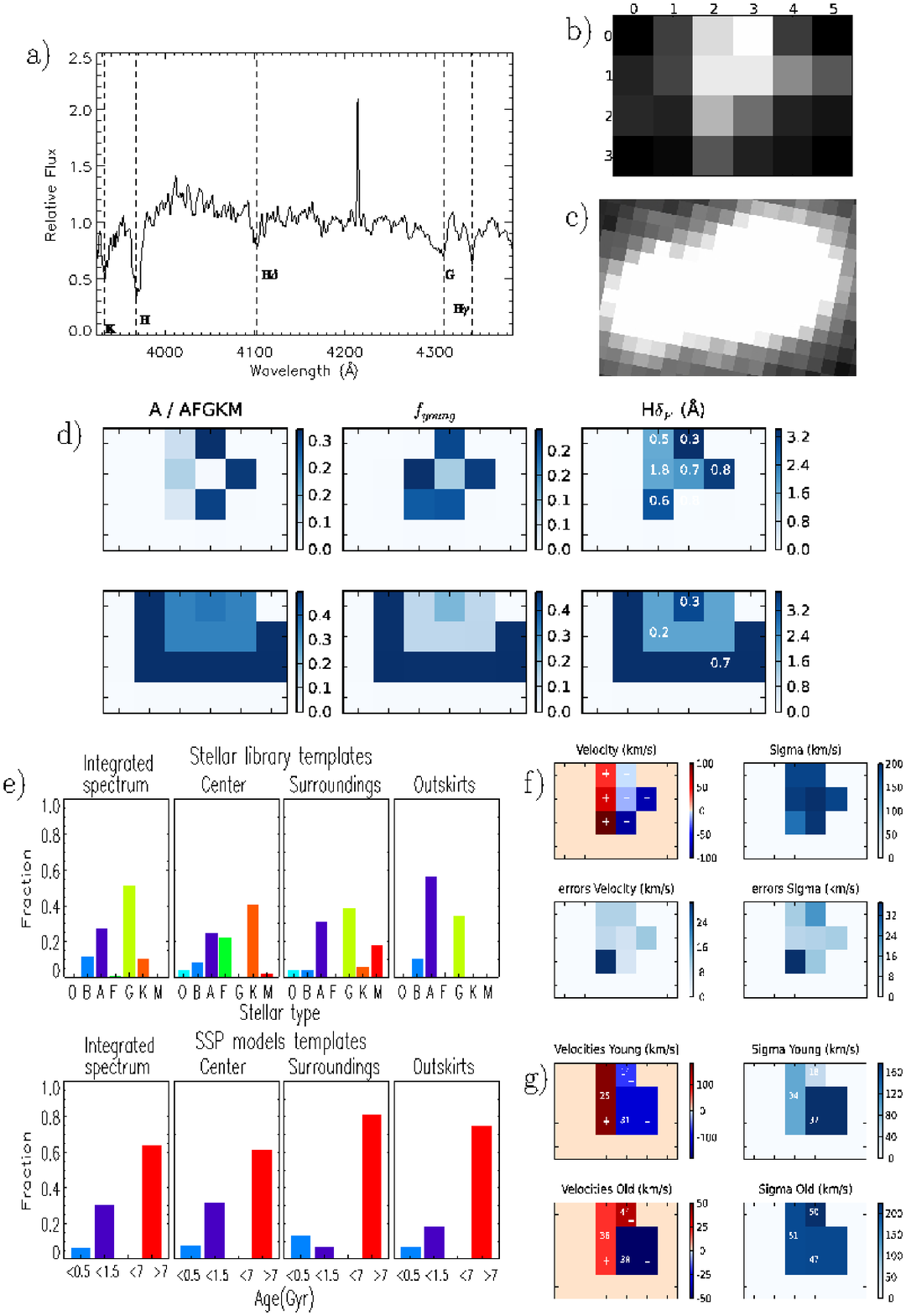}
\end{center}
\end{figure*}

\begin{figure*}
\begin{center}
\LARGE{CN858}
\hspace{0cm}
\vspace{0.2cm}
\includegraphics[height=0.98\textheight]{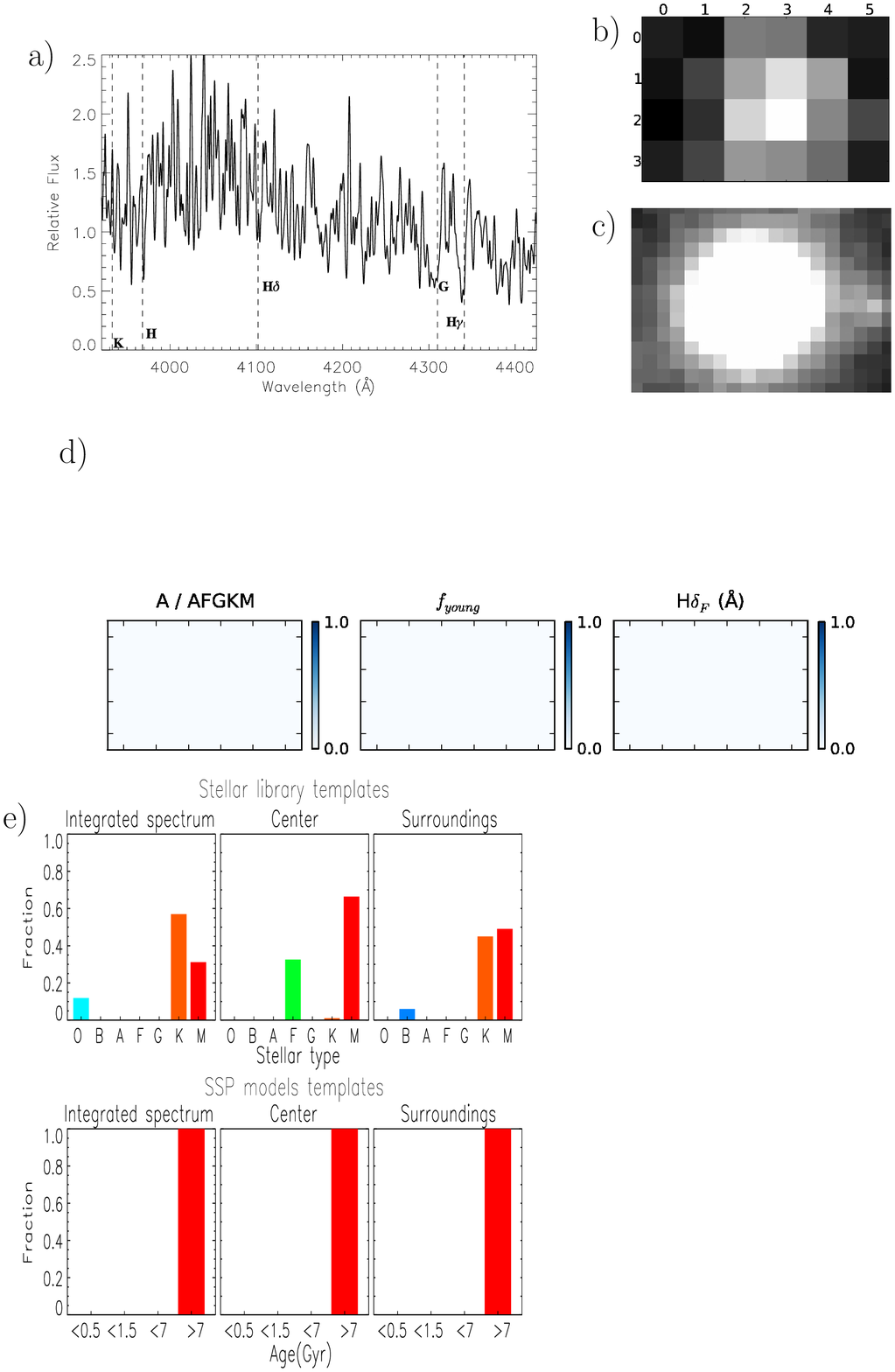}
\end{center}
\end{figure*}

\end{document}